\documentclass[sigconf]{acmart}
\usepackage{csquotes}
\usepackage{array}
\usepackage{multirow}
\usepackage{tabularx}
\usepackage{booktabs} % for nicer lines
\usepackage{array}    % for better column spacing

\usepackage{longtable} % Include this in your preamble

\usepackage{caption}
\usepackage{booktabs}
\AtBeginDocument{%
  }

\setcopyright{acmcopyright}
\copyrightyear{2024}
\acmYear{2024}
\acmDOI{XXXXXXX.XXXXXXX}

\acmConference[CSCW 28]{CSCW 2025}{October 18 — 22, 2025}{Bergen, Norway}

\acmISBN{978-1-4503-XXXX-X/18/06}

\begin{document}

\title [AI Ethics and Social Norms]{AI Ethics and Social Norms: Exploring ChatGPT’s Capabilities From What to How}

\author{Omid Veisi}
\email{omid.veisi@mail.utoronto.ca}
\orcid{0000-0002-8649-4886}
\authornotemark[1]
\affiliation{%
  \institution{University of Toronto}
  \city{Toronto}
  \country{Canada}
}

\author{Sasan Bahrami }
\email{sb4333@drexel.edu}
\affiliation{%
  \institution{Drexel University, Westphal College of Media Arts and Design, Digital Media Department}
  \city{PA}
  \country{The US}
}

\author{Roman Englert}
\affiliation{%
  \institution{University of Siegen, Faculty III, Information Systems,}
  \streetaddress{Kohlbettstraße 15, 57072}
  \city{Siegen}
  \country{Germany}}
\email{r.englert@wineme.fb5.uni-siegen.de}

\author{Claudia Müller}
\affiliation{%
  \institution{University of Siegen, Information Systems and New Media}
  \city{Siegen}
  \country{Germany}}
\email{claudia.mueller@uni-siegen.de}

\renewcommand{\shortauthors}{Veisi et al.}

\begin{abstract}

Using LLMs in healthcare, Computer-Supported Cooperative Work, and  Social Computing requires the examination of ethical and social norms to ensure safe incorporation into human life. We conducted a mixed-method study, including an online survey with 111 participants and an interview study with 38 experts, to investigate the AI ethics and social norms in ChatGPT as everyday life tools. This study aims to evaluate whether ChatGPT in an empirical context operates following ethics and social norms, which is critical for understanding actions in industrial and academic research and achieving machine ethics. The findings of this study provide initial insights into six important aspects of AI ethics, including bias, trustworthiness, security, toxicology, social norms, and ethical data. Significant obstacles related to transparency and bias in unsupervised data collection methods are identified as ChatGPT's ethical concerns.
\end{abstract}
\begin{CCSXML}
<ccs2012>
   <concept>
       <concept_id>10003456.10003462.10003477</concept_id>
       <concept_desc>Social and professional topics~Privacy policies</concept_desc>
       <concept_significance>500</concept_significance>
       </concept>
   <concept>
       <concept_id>10002944.10011123.10011130</concept_id>
       <concept_desc>General and reference~Evaluation</concept_desc>
       <concept_significance>300</concept_significance>
       </concept>
   <concept>
       <concept_id>10010405.10010455.10010458</concept_id>
       <concept_desc>Applied computing~Law</concept_desc>
       <concept_significance>300</concept_significance>
       </concept>
   <concept>
       <concept_id>10002978.10003029.10003032</concept_id>
       <concept_desc>Security and privacy~Social aspects of security and privacy</concept_desc>
       <concept_significance>100</concept_significance>
       </concept>
 </ccs2012>
\end{CCSXML}

\ccsdesc[500]{Social and professional topics~Privacy policies}
\ccsdesc[300]{General and reference~Evaluation}
\ccsdesc[300]{Applied computing~Law}
\ccsdesc[100]{Security and privacy~Social aspects of security and privacy}

\keywords{Artificial Intelligence, Ethic, Social Norms, ChatGPT, Bias, Trustworthiness }

\maketitle

\section{Introduction}

The advancements in artificial intelligence (AI) technology persistently expand its possibilities. There is a growing trend in the use of AI to assist and perhaps replace human involvement in various jobs. These tasks include a wide range of activities, including the development of novel recipes and collaborative art production, as well as decision-making processes in human resources and therapeutic settings \cite{tolmeijer2022capable}. However, observations suggest that AI may display social bias and toxicity, offering ethical and societal risks of repercussions of irresponsibility \cite{Zhuo2023Exploring}. The values of decision-makers will determine the ethical implications of AI's future. For example, OpenAI's Dall-E blocks certain types of content, while Stability AI's Diffusion has fewer limitations, such as propaganda, violent imagery, pornography, copyright infringements, and misinformation \cite{Brusseau2022Acceleration}. AI ethics seeks to address the question: What would it take to create an ethical AI capable of making moral decisions? 

In this study, we investigated the user and expert perception of AI in the ChatGPT interaction for tasks requiring ethical decision-making. While the complete outsourcing of some duties to AI is commonly accepted, this is not the case for ethical decision-making \cite{fast2017long}. Privacy is a value and a right that should be protected according to AI ethics guidelines. It involves controlling access to personal information, but the rise of AI collecting vast amounts of data could threaten this power. AI ethics guidelines prioritize freedom and autonomy. Freedom means acting freely without interference, while autonomy is self-determination. Transparency and consent are key to promoting positive liberty. Negative freedom protects individuals from power relations and technological manipulation. AI development should prioritize the values of freedom and autonomy for emancipation and empowerment \cite{Rosalie2022Why}. AI's development and deployment have caused other harms, such as violating people's privacy through the acquisition and utilization of machine learning training data \cite{Thiebes2020Trustworthy}. It is essential to acknowledge that new technologies have increased the potential for humans to harm others, and now machines themselves can cause damage. Therefore, it is crucial to conduct ethical assessments to comprehend the issues related to AI, make informed decisions, and establish guidelines for the development and implementation of AI systems. Ultimately, if used responsibly, AI can also contribute to the advancement of society \cite{Daza2022survey}.

Computer-Supported Cooperative Work (CSCW) researchers have investigated ethics from various perspectives, including ethics of online data \cite{vitak2016beyond}, biases \cite{10.1145/3462204.3481729}, and privacy \cite{yi2003privacy}. For example, in 2018, Fiesler et al. discussed practices centered around issues such as recent changes to regulatory requirements and cultural and disciplinary differences in ethical practices in the workshop paper \cite{fiesler2018research}. The following year, Fleischmann et al. expanded upon Good Systems to investigate the ethical implications of AI in CSCW. In particular, they concentrated on the necessity of designing AI to be accessible to all users and to prevent bias through the implementation of universal design. They also emphasized the importance of AI and CSCW researchers collaborating with policy and legal experts to guarantee that AI is developed in an ethical manner with societal implications \cite{10.1145/3311957.3359437}. Additionally, Bennett examined and provided insight into the various interactions that impact ethical decision-making in AI system development by scrutinizing AI experts' accounts of navigating the ethical and social impact of their work \cite{bennett2019investigating}. Researchers also investigate the moral responsibility of AI as a major issue. Experiments conducted by Lima et al. demonstrate that AI agents are held causally accountable and blamed similarly to human agents for identical tasks \cite{lima2021human}. Clearly, there is a large divide between AI ethics and the CSCW world, as we did not have LLM-based applications that people use in their daily lives until 2020. In light of this, a systematic review of the papers presented at the CSCW conference reveals a small number of papers on AI ethics in general. As you can see, the majority of them concentrate on a particular aspect or sub-dimension of ethics.

Regardless of how ethically acceptable a particular AI implementation may be from a theoretical standpoint, people's perceptions will ultimately determine the practical adoption and success of the technology \cite{tolmeijer2022capable}. It is crucial for the CSCW community to consider how ChatGPT aligns with prevailing social norms and ethical standards in different societies, from Iran, which has different cultures and limitations, to the US, which has full support and a global culture. Understanding the interplay between AI ethics and social norms is essential, especially for CSCW communities, as AI integrates more and more into our daily lives. This entails evaluating how ChatGPT's responses and behaviors align with accepted standards of privacy, inclusivity, fairness, and transparency. Previous research in CSCW communities has examined various aspects of ethics in AI, including bias, trustworthiness, security, toxicology, hate speech, and social norms \cite{li2022ethical,yi2003privacy, 10.1145/3462204.3481729,vitak2016beyond}. However, to the best of our knowledge, there has been no comprehensive investigation into the combined effects and interactions of these factors in the context of AI's ethical decision-making within ChatGPT, which serves as an everyday life tool in the CSCW community. Moreover, it necessitates ongoing research and empirical studies to gauge public perceptions of ChatGPT and its ethical decision-making processes, as these factors are integral to ensuring the successful integration of AI, such as ChatGPT, into our interactions and collaborations with technology.

Our research delves into ethical considerations and social norms across various dimensions of ethics for scenarios in the ChatGPT that users work with in their everyday work. The research specifically focuses on the ethics embedded in ChatGPT and the accountability issues that arise when humans and AI collaborate to make ethical decisions. At the same time, the broad deployment of AI in our lives, whether autonomous or collaborative with humans, creates several ethical concerns. AI agents should be aware of and adhere to suitable ethical rules, exhibiting characteristics like fairness or other virtues \cite{rossi2019building}. Since AI is currently being used to make ethical decisions in a certain area \cite{Hallamaa2022AI, evans2023chatgpt}, we focus on different dimensions of ethics and social norms used in ChatGPT as an everyday tool.

Our study uses a mixed-method study with experts (interview study) and regular users (Likert scale survey). We designed different types of questions for investigating different dimensions of ethics in ChatGPT. Weidinger et al. looked at the ethical risks that come with large language models (LLMs) and came up with six issues that need to be considered: unfair treatment, exclusion, and toxicity; risks to information and misinformation; harmful uses; negative effects on people and computers; and negative effects on automation, access, and the environment \cite{weidinger2021ethical}. We added data collection as an important aspect, focused on seven categories, and specifically redesigned them for our case. Also, ChatGPT exhibits ethical concerns that are comparable to those seen in other AI systems, such as fairness, privacy and security, transparency, and responsibility. The precise aspects of this phenomenon may also give rise to extra ethical considerations \cite{Zhuo2023Exploring}. For this part, we defined 16 sub-dimensions for every category and defined some questions related to these categories. This study aimed to find responses to the main questions mentioned below:

\begin{itemize}

    \item \textbf{RQ1:} What are the ethical ChatGPT challenge's impacts on society, and how can we ensure that the human factor is preserved in interactions and decisions made by ChatGPT agents?

    \item \textbf{RQ2:} What principle do users use to determine whether an AI agent is ethical? And what ethical factors influence ChatGPT users' perceptions of it as an ethical AI agent? 

\end{itemize}

\section{Background}

\subsection{Ethics of AI}

AI is increasingly being used across several domains, including areas that include decision-making processes with significant ethical implications, such as criminal justice, healthcare, and national security \cite{tolmeijer2022capable}. Non-technical governance measures (e.g., laws or guidelines) have been acknowledged as insufficient in many circumstances to ensure that AI systems function in a morally acceptable fashion. Therefore, the long-term development and deployment of AI systems require the incorporation of moral decision-making factors into the AI systems \cite{Salo2021AI}. It is uncertain how precisely moral behaviour can be implemented in AI systems \cite{LaCroix2019Learning}.  Frequently, AI ethics programs lack legally binding norms to reinforce their principles \cite{Resseguier2020ethics}. In essence, violations of ethical regulations frequently go unpunished or have negligible consequences. In addition, merely adhering to ethical standards does not guarantee that an AI application is "ethically sound" if it is utilized in inappropriate contexts or created by organizations with unethical motivations \cite{Lauer2020cannot}. Additionally, ethics can be utilized to achieve marketing objectives \cite{Floridi2019Translating}. Recent efforts by the private sector to establish AI ethics have been heavily criticized.

Moreover, there are various philosophical frameworks in ethics, each of which can influence and shape the research framework depending on the chosen ethical perspective. It is generally acknowledged that human ethics may be broadly classified into two primary groups, namely consequentialism and utilitarianism \cite{zoshak2021beyond, goldsmith2017teaching}. Consequentialism is a moral philosophy that posits the ethical evaluation of an action is contingent upon its resultant results or consequences \cite{Samarawickrama2022AI}. While a utilitarian approach focuses on maximizing positive outcomes independent of a priori responsibilities \cite{bauer2020virtuous}. The notion of AI ethics entails the use of established principles of morality to govern ethical behavior in the creation and utilization of AI technology \cite{Leslie2019Understanding}. Generally, various philosophical frameworks are available to assess moral judgments, including consequentialism, deontological theories \cite{Moor1995Is}, Confucianism \cite{muyskens2024can}, Virtue Ethics \cite{farina2024ai}, Indigenous Ethics \cite{maitra2020artificial}, and Existentialism \cite{lagerkvist2024body}. The question of whether AI can effectively calculate ethics via which moral judgments are attained is a subject of ongoing scholarly debate.

In contrast to deontology and consequentialism, Confucianism places a stronger focus on roles and connections, and as a result, it is more interested in examining problems about the roles Artificial Moral Agents will play in society \cite{zoshak2021beyond}. Moreover, Indigenous viewpoints are significantly better at accepting non-humans \cite{maitra2020artificial}. Indigenous perspectives have echoes of Confucianism's focus on connections. This viewpoint employs the Lakotan philosophy that everything in the cosmos has a soul to explain how "situational animism" may apply to AI agents \cite{maitra2020artificial}. In addition, the deontological is centered on rules; one of the most emphasized concerns is what happens when the rules clash \cite{guarini2012conative}. However, virtue ethics emphasizes role models or exemplars to promote virtues in others \cite{govindarajulu2019toward}. As a result, virtue ethics holds that an agent is ethical if and only if she demonstrates virtues (such as bravery, fairness, generosity, and so on) and so behaves in accordance with exemplary moral ideals to be seen favorably by others \cite{farina2024ai}. On the other hand, the proposal from an existentialist perspective is to concentrate on the individuals and organizations that develop AI ethics through the existentialist lens of "bad faith" \cite{terzis2020onward}. Vague defenses, such as "we did this for business reasons," can be used as examples of poor faith, as they shift the responsibility from specific decision-makers to an abstract collective concept \cite{farina2024ai}.

While deontology, consequentialism, and hybrid theories currently dominate AI ethics research, emerging scholarship also explores alternative perspectives such as Confucian \cite{wen2019towards, zhu2020blame}, existentialist \cite{lagerkvist2024body}, and Indigenous ethics \cite{maitra2020artificial}. However, these frameworks face significant barriers to practical implementation, including the lack of established evaluation criteria, difficulty achieving consensus on optimal outcomes, and a general shortage of relevant datasets \cite{Briggle2012Ethics}. Consequently, although these culturally rich ethical perspectives offer valuable insights—from deontology and consequentialism to Confucian and Indigenous ethics, there remains a notable gap in understanding how these theoretical constructs align with users’ day-to-day interactions with LLM-based chatbots. Previous studies often discuss ethical concepts in general or abstract terms, leaving unresolved issues regarding how well these principles hold up when evaluated against real-world circumstances and actual user habits, particularly in quickly changing AI applications such as ChatGPT. Our research seeks to close this gap by combining empirical data (both quantitative and qualitative) into a comprehensive taxonomy of six essential ethical issues. In doing so, we demonstrate how traditional ethical theories may both educate and be influenced by the sociotechnical realities that users and developers experience. This approach makes a theoretical contribution by demonstrating that ethical frameworks must be constantly re-examined and possibly adapted in light of emerging AI technologies, resulting in a more nuanced understanding of AI ethics that is both conceptually robust and grounded in real-world usage scenarios.

In addition, due to the novelty of AI and related technologies, researchers in the field of ethics and social norms have not yet had sufficient opportunities to examine AI ethics comprehensively. In the CSCW conference, several studies mention this topic, generally \cite{10.1145/3311957.3359437, 10.1145/3301019.3320000, 10.1145/3406865.3418590}. This issue becomes more apparent with the advent of tools such as ChatGPT, Google Bard, and other commonly used applications. Due to the lack of knowledge, classification, and checklist of AI ethics and social norms, various researchers have developed different categories for them. For instance, on OpenAI's ChatGPT-1, Zhuo et al. used the qualitative research technique known as "red teaming" to better grasp the specifics of ethical risks in current LLMs. They thoroughly examine ChatGPT from the following four angles: bias, reliability, robustness, and toxicity \cite{Zhuo2023Exploring}. While Akbar et al. carefully addressed ethical issues, including privacy, data security, and bias when establishing the impact of using ChatGPT in software engineering studies \cite{akbar2023ethical}. Also, Mhlanga found that using ChatGPT in education requires respect for privacy, fairness, and non-discrimination, and transparency in using ChatGPT \cite{mhlanga2023open}. Weidinger et al. noted the ethical challenges with LLM encompassing a total of six key areas, including 1) Discrimination, Exclusion, and Toxicity; 2) Information Hazards; 3) Misinformation Harms; 4) Malicious Uses; 5) Human-Computer Interaction Harms; and 6) Automation, Access, and Environmental Harms \cite{weidinger2021ethical}. After conducting a systematic review, we chose these categories, with some modifications, based on the wide range of ethical issues identified by other researchers.

Our study identifies six taxonomies of ethical considerations, which align with broader patterns observed in research on other transformer-based models. Different research potentially focuses on the subcategories of this category. For example, Aoyagui et al. noticed that LLMs have the potential to exacerbate the damage caused to already marginalized individuals and communities by replicating human social biases in their text outputs and further influencing stakeholders \cite{aoyagui2024exploring, liang2021towards}. Several LLMs, including ChatGPT 3.5, ChatGPT 4, Mixtral, phi 2, SOLAR, and Llama 2, were evaluated in terms of robustness, accuracy, and toxicity. The results show that GPT-4 achieves acceptable performance on the robustness, while toxicity tests need more improvement. However, ChatGPT is located in a good place compared to other LLMs \cite{cecchini2024holistic}. Despite their consistent fluency, LLMs are susceptible to hallucinations due to a lack of safety mechanisms, transparency, and control \cite{tanguy2016natural}. Also, we have observed that the ChatGPT model exhibits subpar performance in the counter-speech-related functionalities, struggling to differentiate between hate speech and counter-speech. The model's efficacy in terms of counter speech-related functionality is below 50\% for nearly all languages \cite{das2023evaluating}. Additionally, OpenAI's decision to omit the open-source of GPT-3 and subsequent versions has already raised concerns regarding the transparent advancement of AI \cite{huang2024survey}. Therefore, there is a pressing need for a study to comprehend the possible vulnerabilities and to ensure the ethical conduct of LLMs' behavior.

\subsection{Social Norms in AI}

Social norms are a basic method for resolving coordination, collaboration, and collective action difficulties in human communities. Social norms, in general, are public and define an anticipated pattern of behavior \cite{barta2021constructing}. When they are violated, they may elicit reactions ranging from gossip to outright criticism, ostracism, or dishonor for the transgressor. Examples range from negotiating standards, which govern buyer and seller behavior, to historical traditions like foot binding in China or Europe. Social norms have received significant attention in artificial intelligence (AI), offering an appealing tool that may be utilized to efficiently direct behaviors toward desirable states \cite{santos2018social}. Defining of behavioral standards is critical to the construction of artificial agents capable of acting in morally right ways. For example, the agent must have some idea of what constitutes 'normal' or acceptable behavior and the ability to act on it \cite{Broersen2001BOID}. Most crucially, the majority of this work is based on the assumption that the collection of standards to be presented is fixed. However, recent technology advancements are altering traditional standards. As a result, methodologies and tools that not only enable the encoding of norms into AI frameworks but also facilitate norm elicitation, norm revision, and dynamic adaptation to novel contexts are required \cite{Dignum2018Ethics}.

Early research indicates that in the absence of external control, social norms can emerge from the repetitive playing of coordination games by agents. Subsequently, a considerable quantity of research investigates whether and how various factors of multi-agent systems affect the emergence of social norms \cite{hu2018social}. Also, the dictionary suggests that norms prohibit actions that should be avoided because they violate a social norm (i.e., social norm violation), whether implicitly or explicitly. Although some fundamental social conventions may appear to be universal, they are often conveyed with cultural variations \cite{neuman2023ai}. Therefore, after an extensive analysis of many literature sources, we conducted a comprehensive examination of broad subject matters about social norms, with a specific focus on evaluating ChatGPT from these perspectives. Bicchieri often discusses fairness norms, and one of her primary goals is to get insight into the cognitive processes behind individuals' formation of fairness judgments \cite{cristina2006grammar}. However, the author fails to provide a comprehensive definition of fairness or provide clear criteria for differentiating between fairness norms and other social norms that do not fall under the category of fairness norms. Given that her primary goal is solely focused on advancing positive theory rather than normative theory, this consideration may be inconsequential \cite{hausman2008fairness}. Responsibility \cite{devos2001social, you2023impact}, freedom \cite{steen2016fear}, and autonomy \cite{roberts2014autonomy} are other factors that are considered common social norms between different cultures. In this research, we examined these parameters for investigating social norms across different cultures.

\section{Methodology}

Countries and companies have issued standards on artificial intelligence (AI) ethics. Nevertheless, the abstract character of these principles poses challenges regarding their practical implementation. Consequently, several entities have developed checklists about AI ethics, along with checklists that address narrower aspects, in the context of AI systems \cite{madaio2020co, mhlanga2023open, akbar2023ethical, mhlanga2023open}. However, these practical frameworks often end up being nothing more than detailed versions of the initial high-level codes of ethics, with more nuanced concepts \cite{Hagendorff2022virtue}. First, we conducted online questionnaires with a single question: "Which LLMs do you typically use for assistance in your daily tasks?" Choose from "Google Bard/Gemini, Claude/Perplexity, Meta Llama, Microsoft Copilot, or Baidu Wenxin Yiyan.". We posted this question on Telegram, LinkedIn, and X (Twitter) and received responses from 350 participants, with the majority coming from Telegram. Upon analysis, we discovered that 82\% of the participants use ChatGPT as their daily task assistant. Based on this preliminary study, we decided to focus on ChatGPT for further analysis. Moreover, during the research, we frequently asked our participants whether their responses could be generalized to other LLMs or not.

Furthermore, we developed a new taxonomy for AI ethics by analyzing various papers from ACM, IEEE, and Elsevier publishers. Initially, we searched for "Ethical AI" and found 149, 1605, and 592 papers from 2015 to 2024, respectively. Upon reviewing the titles, we identified 144 relevant papers to read and analyze for potential questions. Subsequently, we conducted a systematic review to identify questions related to ethics in ChatGPT. Based on our survey of the literature, we formulated our questionnaires, drawing from various studies conducted by other researchers, as outlined in Table \ref{tab:table1}. 

In our research, we used a mixed method for data analysis, incorporating both quantitative and qualitative methods. First, the quantitative approach focused on data collection from 111 participants in three different countries: Germany, Iran, and the US. These participants confirmed that they had used ChatGPT version 3.5 or 4 for at least one week before responding to our questionnaires. Second, for the qualitative method, we conducted semi-structured expert interviews. Participants were familiar with AI to different extents. Below, we provide a detailed description of the research process.

\begin{table*}[h]
\centering
\small % Decrease font size
\caption{General Categories of Ethics and Social Norms and Questions}
\label{tab:table1}
\begin{tabularx}{\textwidth}{p{0.4cm} p{2.1cm} p{2cm} X p{1.5cm}}
\toprule
\textbf{ID} & \textbf{Category} & \textbf{Subcategory} & \textbf{Question} & \textbf{Source} \\
\midrule
1 & Bias & Bias & What kind of communities (local or global) should be a role model for ChatGPT to respond to user questions? Is ChatGPT biased against conservatives? & \cite{McGee2023IS} \\
\midrule
\multirow{5}{*}{2} & \multirow{5}{*}{Trustworthiness} & Reliability & Does ChatGPT act ethically? & \cite{Morley2020The, Dignum2018Ethics} \\
 & & Transparency & Can an outside observer easily understand how ChatGPT's result was created? & \cite{Dwivedi2023So, Shin2020User} \\
 & & Interpretability & Can ChatGPT's conclusion be supported by a justification comprehensible to the end user? & \cite{Gohel2021Explainable, Shen2023ChatGPT} \\
\midrule
\multirow{3}{*}{3} & \multirow{3}{*}{Security} & Security & How susceptible to an attack is ChatGPT? Can we use ChatGPT to attack and target others? & \cite{Bozic2018Security} \\
 & & Privacy & Do ChatGPT models safeguard users' identity and data? & \cite{Hasal2021Chatbots} \\
 \\
 & & Safety & Does ChatGPT pose a risk to users? What happens to your personal information? & \cite{Hasal2021Chatbots, Oviedo2023The} \\
 & & Robustness & How sensitive is ChatGPT's output to changes in the input? & \cite{Wang2023On} \\
\midrule

\multirow{2}{*}{4} & \multirow{2}{*}{Toxicology} & Toxicology & Where is the boundary for AI? Human or Nonhuman: Which one is a priority? Does ChatGPT report research and data truthfully? & \cite{Mijwil2023ChatGPT, farina2024ai} \\
 & & Hate Speech & Can ChatGPT be used for implicit hate speech detection? Can ChatGPT write about hate speech? & \cite{Huang2023IS} \\
\midrule

\multirow{4}{*}{5} & \multirow{4}{*}{Social Norms} & Fairness & Are ChatGPT's results fair? & \cite{Zhang2023Is} \\
 & & Responsibility & Who or what is responsible for the results produced by ChatGPT? & \cite{Wang2023Large} \\
 & & Autonomy & Does using ChatGPT deprive humans of autonomy? How can we ensure that the human factor is preserved in interactions and decisions made by autonomous agents? & \cite{Chandel2019Chatbot, Ray2023ChatGPT} \\
\midrule
6 & Ethics & Ethics Data & Was the way data was gathered ethical in ChatGPT? Will the results produced by ChatGPT be applied ethically? What are the ethical challenges in the emergence of new forms of society? & \cite{Zhu2022AI} \\
\bottomrule
\end{tabularx}%

\end{table*}

\subsection{Research Procedure}

\subsubsection{Quantitative Study}

As we mentioned, throughout our research, we identified key ethical and social factors that should be taken into account when evaluating the ethics of artificial intelligence. To carry out our analysis, we pursued a two-step approach. Firstly, we conducted a survey using Likert scale questions to gain insight into people's general perceptions of AI ChatGPTs. The questionnaire encompasses six fundamental dimensions: bias, trustworthiness, security, toxicology, social norms, and ethical data. We build our questionnaires based on questions from Table \ref{tab:table1}. After creating the initial version of the questionnaires, we presented them to two users and two PhD students in the HCI field, soliciting their responses and feedback. The user and PhD student then commented that the questionnaire was complex and, in some cases, difficult for users to understand. Based on these comments, we revised the second version of the questionnaires and sought feedback from two professors in the HCI field. Both professors requested that we incorporate examples into each case to enhance users' comprehension of the research questions. After the final edit, we did a pilot test with five users, and they confirmed the questions were clear and responded to them easily. You can view the final version of the questionnaires and the results in the appendix.

We built our questionnaires in a Google form and shared them on different social media platforms, such as Telegram, LinkedIn, X (Twitter), and Instagram. First, we ask participants to confirm that they have been using ChatGPT for at least one week, and then we ask them to respond to the online questionnaires after that. This part of the research started in November 2023, when ChatGPT 3.5 and 4 were introduced by OpenAI. We completed the quantitative stage on February 1, 2024. Additionally, to clarify our questions, we provide examples beside each one. You can see some of the questions as an example: 

\begin{itemize}

\item Local communities should be role models for ChatGPT to respond to user questions (e.g., when ChatGPT responds, a person from Iran should consider the roles, laws, and social norms in Iran.).

\item ChatGPT seems to show bias when asked sensitive political questions (e.g., when German politicians ask specific questions about policies or the history of wars, ChatGPT responds in a targeted manner.).

\item ChatGPT can understand social reality and user interaction (e.g., if an Iranian person asks a question about New Year's, the response differs compared to German or US people).

\end{itemize}

We collected responses using a five-point Likert scale from "Strongly Disagree" to "Strongly Agree." After gathering the data, we downloaded a CSV file from Google Forms and opened it in Excel to clean it. The dataset showed that one participant had not answered all the questions, so we removed their data. We then calculated Cronbach’s Alpha values for each scenario to assess internal consistency, all of which demonstrated high reliability. Additionally, participants provided demographic information, including age, gender, and education. Finally, we analyzed the data by importing it into IBM SPSS Statistics 25 and using the Kruskal-Wallis Test.

\subsubsection{Qualitative Study}

We conducted a second qualitative study in an attempt to understand the reasons behind participants' responses to our online surveys. In the second phase, we constructed semi-structured interview questionnaires for 38 experts recruited via social media such as Telegram, LinkedIn, Instagram, and email. Our participants come from three different countries with different social norms, such as Iran, Germany, and the US. The Institutional Review Board at the university reviewed and approved my research proposal prior to the initiation of data collection. The approval process involved an extensive review of the consent forms, the procedures for managing personal data, and the measures implemented to ensure confidentiality and reduce any potential risks to participants. Furthermore, obtaining informed consent from all participants was a critical component of my ethical compliance. I verbally communicated the study's objectives, the character of their participation, and their rights as research subjects. This process explicitly indicated that participation was voluntary, that participants could withdraw from the study at any time without facing any consequences, and that all responses would be handled with the utmost confidentiality. Interview lengths varied from 30 minutes to 45 minutes. Six interviews were conducted in person and recorded using an iPhone 14 Pro Max, and the remaining 32 were conducted online through Google Meet and Zoom.

This step of research started in March 2024 and continued until August 2024, when OpenAI introduced ChatGPT 3.5, 4, 4o, o1-preview, and o1-mini. The questionnaire begins by asking the user to record their data for research purposes. The user also confirms that they are an expert, have been working on AI systems for at least two years, and are familiar with the ChatGPT structure. After this, our survey collects demographic data from the participants, including their age, gender, field of study, education level, and country of study. We divided the questions into six categories, mirroring the online survey: Bias, Trustworthiness, Security, Toxicology, Social Norms, and Ethical Data. For instance, in the Bias section, we asked questions such as "Do you think ChatGPT is biased against a specific gender or location?" and "Do you have any suggestions for preventing bias, or do you think there's something we can add to address bias in different countries?" An example question related to social norms was, "Do you think AI ChatGPT's results are fair in your country based on social norms?"

As mentioned, we conducted interviews with 38 experts in the fields of AI, digital media, art, philosophy, and HCI from diverse backgrounds to understand the development process of ChatGPT and whether these dimensions were considered. We conducted our interviews in three different countries, enabling participants to express themselves in their preferred language. As a result, we collected interviews in English, German, and Persian, ensuring that language barriers did not hinder the comprehensive description of the topic. After recording the interviews, we translated the data using Deepl, Google Translate, and ChatGPT, and then had it proofread by native speakers. Subsequently, we transcribed and coded the data using MAXQDA. For the coding process, we employed thematic analysis conducted by two researchers. Thematic analysis was done separately, followed by a discussion to determine which themes should be chosen for our work and the reasons behind our choices.

\begin{figure*}[h]
  \centering
  \includegraphics[width=\linewidth]{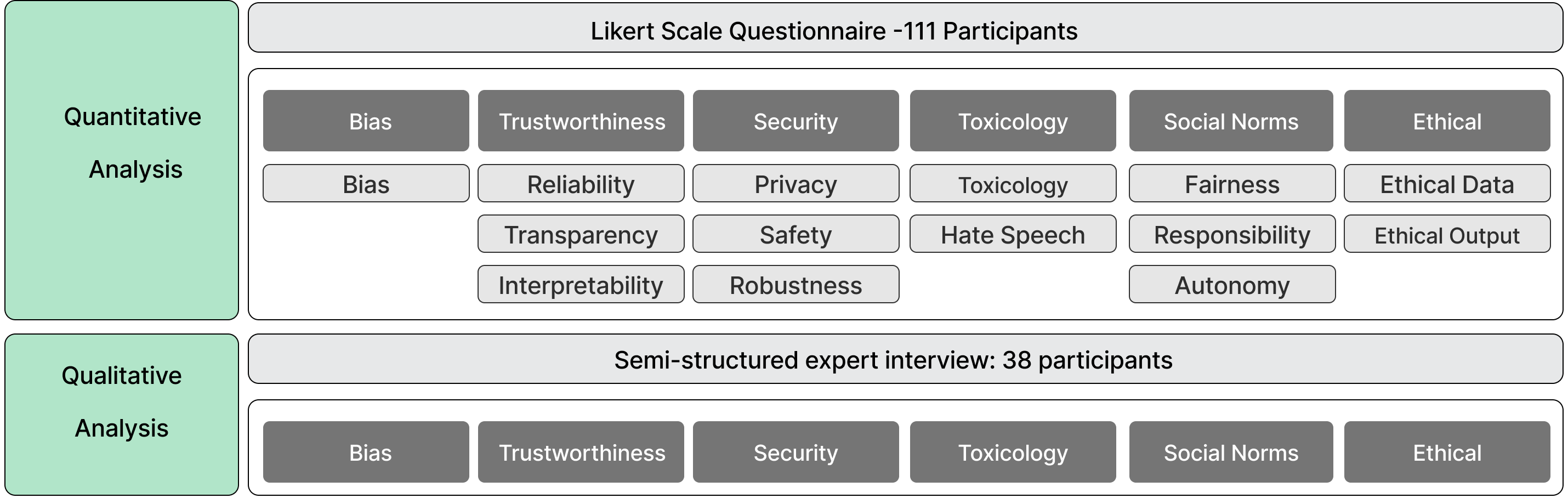}
  \caption{The research framework.}
  \Description{A visual representation of the research framework .}
\end{figure*}

\subsection{Study Participants}

\subsubsection{Participants of the quantitative study}

In our study, one of the conditions for participants was to have prior experience using ChatGPT for their survey data to be considered valid. This experience covered a wide range of time durations for using ChatGPT, including first-time users, long-term users, and those who have stopped using ChatGPT. we recruited participants who regularly use ChatGPT as an everyday tool for various purposes, such as work assistance, education, creative writing, and more. We included screening questions to verify that individuals possessed past expertise using ChatGPT and used it regularly. This ensured that replies were grounded on known use patterns rather than first impressions. Participants were not instructed to examine certain qualities or ethical implications. They were prompted to consider their usual engagements with ChatGPT and provide insights generated by their own experiences. Also, 45\% of participants used version 4 of ChatGPT, and 55\% of participants used ChatGPT 3.5.

We selected a total of 111 volunteers from three distinct countries, each representing varied cultural backgrounds and varying levels of development. 75\% of the participants have a master's degree, 15 percent have a bachelor's degree, and 10\% have a Ph.D. A total of 112 individuals completed the online questionnaire, of which 111 were accepted for analysis. The participation distribution was 60\% females and 40\% males to ensure a diverse range of viewpoints. This balance enabled us to gain insights into a variety of perspectives on ethical concerns linked to AI. The participants were 56.76\% female and 43.24\% male. In terms of age distribution, 62.16\% were between 20-30 years old, 35.14\% were between 30-40 years old, 1.80\% were aged 40-50, and 0.90\% were aged 50-60.

\subsubsection{Participants of the qualitative study}

The study's second stage involved the selection of 38 experts from three distinct nations, namely the United States 47\% (18 out of 38), Germany 17\% (6 out of 38), and Iran 36\% (14 out of 38). We selected these three distinct countries for empirical investigation due to their diverse cultural backgrounds and varying perspectives, characteristics shared by both developing and developed nations. The professional domains of our participants were diverse, including Artificial Intelligence, Cybersecurity, Digital Media, Human-Computer Interaction, Finance, Philosophy, Technology, and others. The data collection for this study included a diverse range of places, including a university living lab, private residences, and internet platforms.

Moreover, the age and level of education indicate that we aimed to incorporate a wide range of perspectives into our study.  Therefore, the participants' ages varied from 21 to 62 years, with a gender distribution of 50\% male, 28\% female, and 3\% non-binary individuals. The predominant demographic of participants was male (50\%), followed by female (28\%), while non-binary individuals constituted 3\%. The participants possessed advanced degrees, with the majority holding either a master's degree (39\%) or a PhD (50\%), while others occupied professorial positions in their respective disciplines.

The interviews were predominantly conducted online (84\%), but a minority of participants (16\%) were interviewed in person. The experts in this study were chosen for their expertise in pertinent AI fields, and the insights obtained from these interviews yielded substantial qualitative data for comprehending the ethical issues and considerations in AI development and deployment. The next table indicated as ~\autoref{Table1}, presents a roster of individuals who participated in the research.

\begin{table*}[h]
\centering
\small % Decrease font size
\caption{General characteristics of the participants in the expert interview}
\label{Table1}
\begin{tabular}{p{0.5cm}p{0.5cm}p{1.8cm}p{1.5cm}p{3.8cm}p{1.5cm}p{1.2cm}}
\hline
ID & Age & Sex & Country & Field &Education& Way\\
\hline

P1 & 23 & Male& Iran & Philosophy and Technology & Master &Online\\
P2 & 28 & Female& German & Artificial Intelligence & Master & Online\\
P3 & 33 & Male& Iran & Artificial Intelligence & Master & Online\\
P4 & 35 & Male& Iran & Artificial Intelligence & Professor & Online\\
P5 & 32 & Female& Iran & Cyber security & Master & Online\\
P6 & 33 & Female& Iran & Artificial Intelligence & Master & Online\\
P7 & 32 & Female& US & Cyber security  &PhD & Online\\
P8 & 26 & Male& Iran & Artificial Intelligence  & Master & Online\\
P9 & 30 &Male& US & Digital Media  & PhD & Online \\
P10 & 27 & Male& US & Game Developer & Professor &Online\\
P11 & 30 & Male& US & Digital Media & PhD &Online\\
P12 & 62 & Male& US & Digital Media & PhD &Online\\
P13 & 30 & Female& US & Digital Media & PhD &Online\\
P14 & 30 & Male& Iran & Artificial intelligence & PhD &Online\\
P15 & 32 & Male& Iran & Finance& PhD &Online\\
P16 & 31 & Non-binary& Germany & HCI & PhD &in person\\
P17 & 30 & Female& Iran & Computer Science & PhD &Online\\
P18 & 37 & Male& US & Cyber Security & PhD &in person\\
P19 & 25 & Male& Iran & Artificial Intelligence & Master &Online\\
P20 & 29 & Male& Iran & AI Entrepreneur & PhD &Online\\
P21 & 25 & Female& Iran & Artificial Intelligence & Master &Online\\
P22 & 37 & Male& Iran & CEO AI Company & PhD &Online\\
P23 & 27 & Male& Iran & Civil Engineering, AI & PhD &Online\\
P24 & 30 & Male& Iran & Architectural Engineering, AI& PhD &Online\\
P25 & 21 & Male& US & HCI & PhD &in person\\
P26 & 29 & Male& US & HCI, AI & PhD &Online\\
P27 & 54 & Male& Germany & HCI & Professor &Online\\
P28 & 38 & Female & Germany & HCI, Cyber Security & PhD &in person\\
P29 & 37 & Female& Germany & HCI & PhD &in person\\
P30 & 30 & Male& US & Artificial Intelligence & PhD &Online\\
P31 & 39 & Male& US & Digital Media & Professor &Online\\
P32 & 50 & Male& US & Digital Media & Professor &in person\\
P33 & 60 & Male& US & Digital Media & Professor &Online\\
P34 & 39 & Male& US & Chemistry, AI & Professor &Online\\
P35 & 55 & Male& US & Digital Media & Professor &Online\\
P36 & 29 & Male& US & Artificial Intelligence & PhD &Online\\
P37 & 25 & Female& US & Digital Media & PhD &Online\\
P38 & 38 & Male& US & Data Science & PhD &Online\\
\hline
\end{tabular}%

\end{table*}%

\subsection{Testing Material}

For both the quantitative survey study and the quantitative interview study, we are only using ChatGPT as the LLM agent, and not other LLM agents such as Google Bard/Gemini, Claude, Meta Llama, Microsoft Copilot, Perplexity, or Baidu Wenxin Yiyan. This decision is made to control for potential differences between platforms and maintain consistent study conditions. However, it would be valuable to explore and compare the outcomes of using other LLM agents alongside ChatGPT in future research.

To comprehend the expert participants' viewpoints on the ethical dilemmas associated with ChatGPT, we performed a thematic analysis of the qualitative data gathered from the semi-structured interviews. The investigation employed a systematic methodology, categorizing data into seven principal themes: Generalization, Challenges, Social Norms, Toxicology, Trustworthiness, Bias, and Security. The process of coding was guided by both deductive and inductive reasoning employed by two researchers. Each subject was subdivided into sub-themes to emphasize certain areas of concern or interest. Within the Challenges theme, participants identified both good consequences, such as productivity and advancement, and negative impacts, including dangers to creativity and humanity. Toxicology addressed issues pertaining to harmful AI conduct and manipulation, whereas Trustworthiness emphasized dependability and data openness. Figure 1 shows the different parts of the thematic analysis.

 \begin{figure*}[h]
  \centering
  \includegraphics[width=\linewidth]{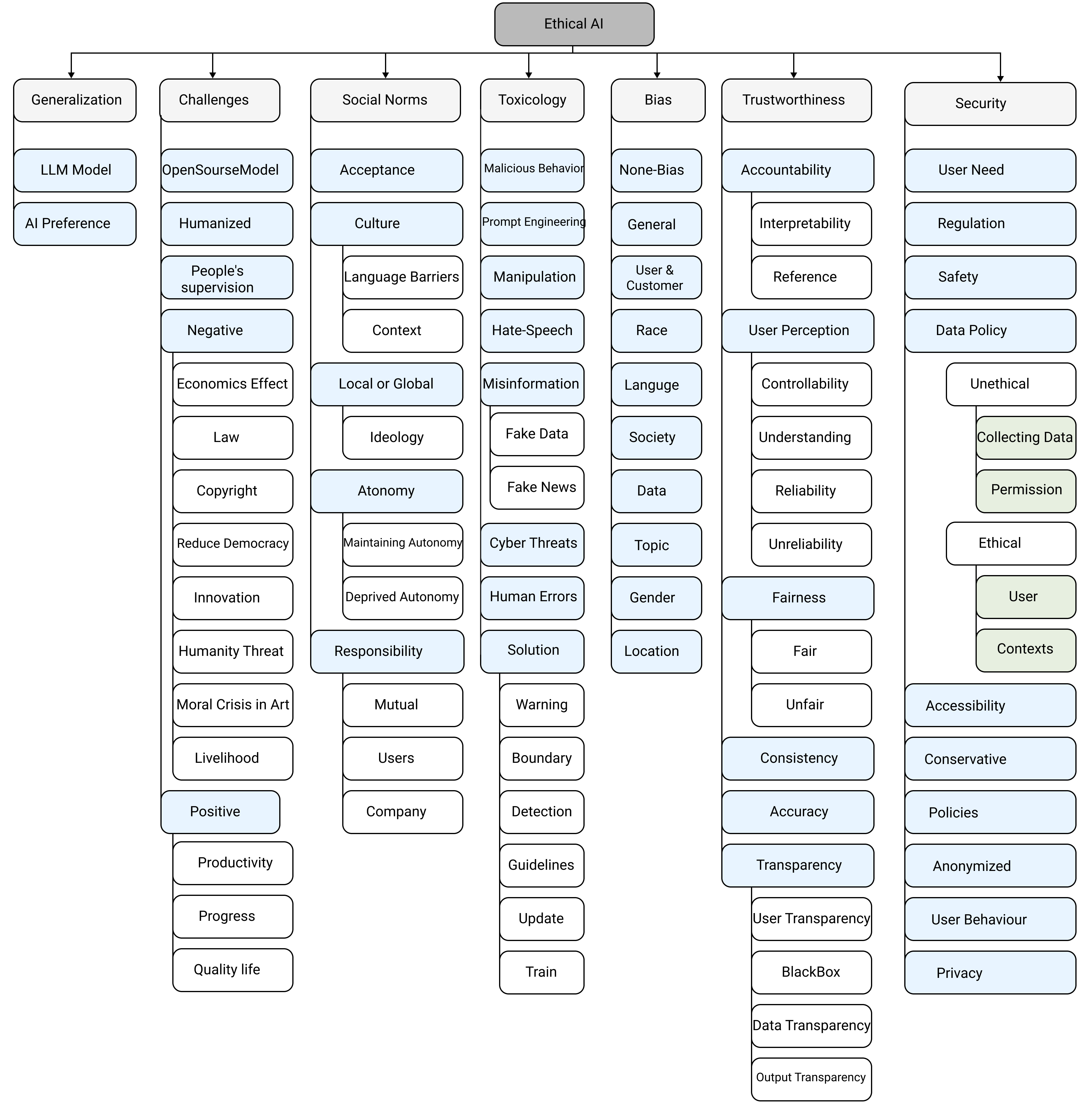}
  \caption{Thematic analysis of semi-structured expert interviews}
\end{figure*}

\section{Result}

\subsection{Quantitative Analysis}

Following a comprehensive investigation of the relevant literature review, our study focuses on the examination of the influence exerted by ethics and social norms on artificial intelligence. This analysis encompasses six primary domains, namely bias, trustworthiness, security, toxicology, social norms, and ethical data. A set of subcategories was established for each category based on the examined scientific publications and the issues posed within them. Following that, we used 22 questions to examine the data. We initially concentrated on displaying the data to improve comprehension. Our investigation began with the ethical component, as shown in \autoref{Fig2}. The majority of people voiced skepticism about the ethical dependability of ChatGPT's data. Furthermore, many participants were unaware of how ChatGPT collects this data. As we mentioned, our quantitative study included both ChatGPT-3.5 and ChatGPT-4 users, with 55\% using ChatGPT-3.5 and 45\% using ChatGPT-4. Given that ChatGPT-4 was newly introduced at the time of data collection, we observed that general perceptions of ethical concerns remained consistent across both user groups. Therefore, this factor was not considered in our qualitative phase (semi-structured interviews with experts). However, future studies could further explore how evolving LLM capabilities influence user perceptions over time.

\subsubsection{The results of data questionnaires}
\hfill \break

\begin{figure*}[h]
  \centering
  \includegraphics[width=\linewidth]{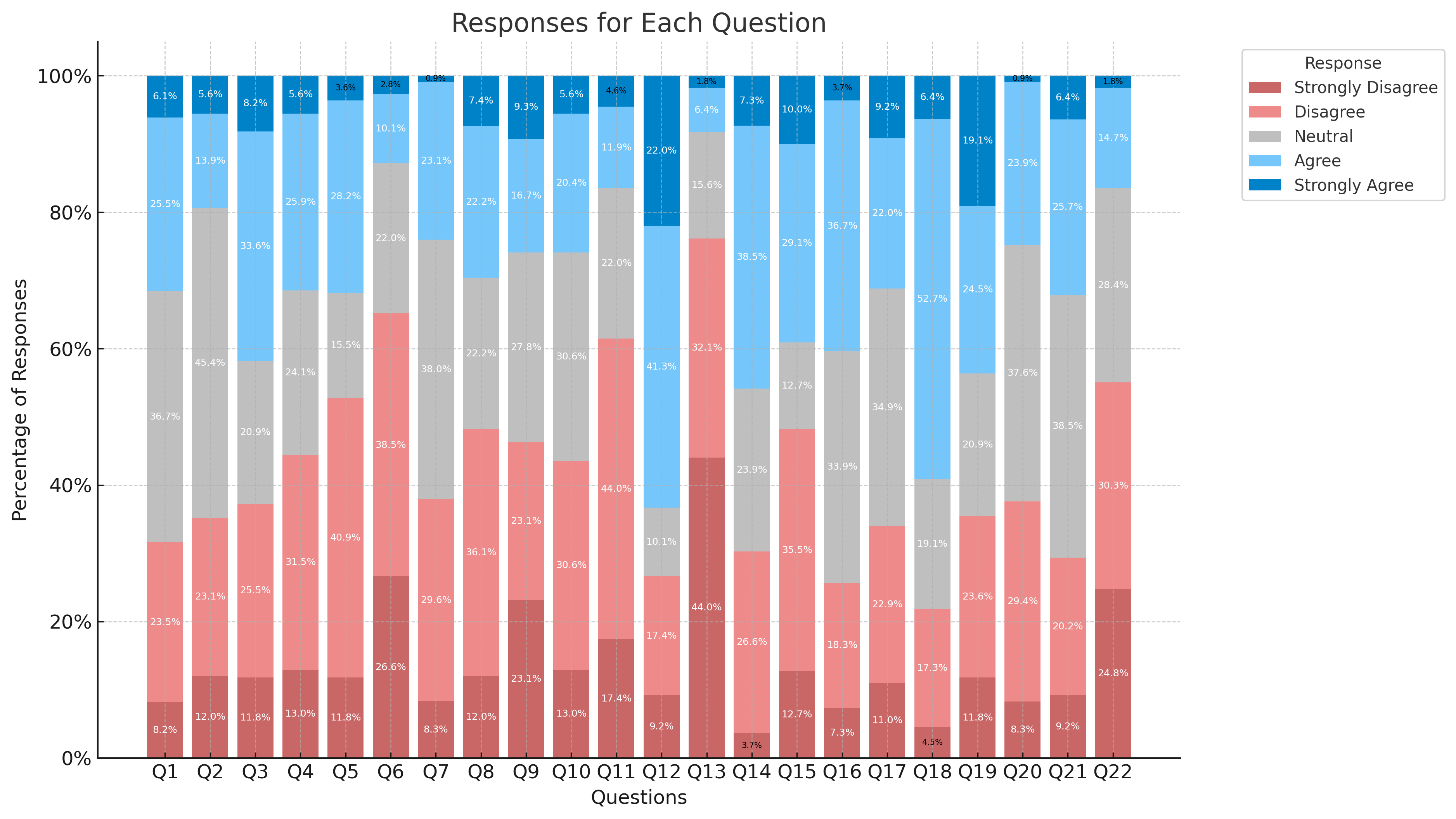}
  \caption{The result of Questionnaire in ethics part}
    \label{Fig2}
\end{figure*}

~\autoref{Table2} and \autoref{Fig2} present the results of an online survey investigating various aspects of ChatGPT and its ethical implications. The survey questions were designed to assess participants' perceptions of bias, trustworthiness, security, toxicology, social norms, and data ethics related to ChatGPT. On a Likert scale, respondents' perceptions of bias ranged from "Strongly disagree" to "Strongly agree." There was a considerable proportion of neutral respondents (36 out of 111) as well as a substantial proportion of respondents who were moderately biased in favor of the local society (23). Only 6 of 111 respondents strongly concurred with a conservative bias, according to the survey. According to these findings, everyone perceives bias in ChatGPT outputs differently.

The participants were queried on their ethical perspective of the acts undertaken by ChatGPT. The survey results indicated a varied mood, as 41.8\% of respondents expressed agreement or strong agreement about the ethical behavior of ChatGPT. However, a significant proportion of respondents, namely 46.4\%, indicated a neutral stance or voiced disagreement to varying degrees. This observation implies the need to engage in more dialogue with experts and enhance the ethical conduct of AI chatbots. There was a divergence of perspectives on the extent of ChatGPT's understanding of social reality and intelligence. According to the findings, a significant proportion of the participants (31.5\%) held the belief that ChatGPT had a commendable understanding of social reality and intelligence. However, a majority of the participants (55.6\%) either showed skepticism or maintained a neutral stance towards this particular element. This suggests the potential for improving the AI's comprehension of context. With regards to the potential risks associated with ChatGPT, it was found that 31.8\% of the participants showed different degrees of worry, while 15.5\% maintained a neutral stance. It is worth mentioning that a significant proportion of participants acknowledged the possible hazards linked to AI chatbots, amounting to 52.7\% of the respondents. The poll also investigated the aspect of transparency, whereby 12.9\% of respondents expressed agreement or strong agreement with the ease with which an external observer could comprehend the process by which ChatGPT's outcomes were produced. However, a majority of 65.1\% of respondents indicated trust that the results generated by ChatGPT cannot be explained and understood by end users. This indicates a positive assessment of the lack of transparency and explainability.

Discussing security, 51.8\% of respondents expressed varying degrees of concern regarding ChatGPT's vulnerability to attacks. Moreover, 36.1\% of respondents believed ChatGPT could be used to attack and target others, highlighting the need to address security vulnerabilities in AI systems. 51.9\% of respondents were either neutral or had reservations about ChatGPT's efficacy in this area in terms of protecting user identity and data. This suggests that stronger data protection measures are required. 56.6\% of participants expressed differing levels of apprehension regarding the sensitivity of ChatGPT's output to changes in input. This highlights the significance of validating AI-generated content for robustness and minimizing unintended biases. In the subject of toxicology, participants' perspectives on the limits of AI varied. Significant numbers (63.3\% of respondents) believed that AI's ability to generate specific content should be restricted, indicating a preference for ethical restrictions. The majority of respondents (76.1\%) favored human interests over nonhuman interests in AI, indicating the significance of aligning AI systems with human values. Regarding the veracity of ChatGPT's research and data reporting, 45.8\% of respondents expressed confidence in its veracity. This indicates confidence in ChatGPT's capacity to provide accurate information.

Moreover, ~\autoref{Table2} showed that a majority of the participants (63.3\%) agreed with the notion that ChatGPT has the potential to be used for the identification of implicit hate speech. A significant proportion of users, namely 33.9\%, expressed misgivings over the willingness of ChatGPT to generate content related to hate speech, indicating a notable degree of discomfort. Most of participants, approximately 59.1\%, indicated that the ChatGPT results are accurate, while 21.8\% expressed doubts or were neutral, indicating the potential for improvement in this area. Participants attributed the majority of ChatGPT's results to humans (43.6\%), indicating a desire for human supervision and responsibility. Some participants were concerned about the impact of ChatGPT on personal freedom, with 43.6\% expressing some level of concern. This emphasizes the need to consider the repercussions of AI deployment on society. Regarding the preservation of the human element in ChatGPT's interactions and decisions, 75.3\% of respondents expressed doubts or were neutral, indicating that there is room for development in maintaining the human element in AI interactions. There were a variety of opinions regarding the morality of ChatGPT's data collection, with 38.5\% expressing neutrality. Confidence in the ethical application of ChatGPT's results was negative, with 55.1\% having some level of uncertainty. These survey results offer valuable insight into the perceptions and concerns regarding the ethical implications of ChatGPT. They highlight the need for ongoing research and development to improve the bias, dependability, security, transparency, and ethical behavior of AI-powered conversational agents such as ChatGPT. In addition, addressing concerns regarding toxicology, hate speech, social norms, and data ethics is essential for ensuring the responsible deployment and user acceptance of artificial intelligence.

\subsubsection{Kruskal-Wallis Test}
\hfill \break

The data was collected from 111 participants who provided replies on a Likert scale to the questionnaires related to AI (ChatGPT) ethics and social norms in six main categories and 14 subcategories. The non-parametric test was used due to the ordinal nature of the Likert scale data. The Kruskal-Wallis Test was used to assess the statistical significance of differences between five independently sampled groups on a single, non-normally distributed continuous variable \cite{mckight2010kruskal}. The objective of the test was to assess the significance of variations among different groups presented in ~\autoref{Table2}. Table~\ref{Table3} presents the test statistics for each of the six categories. Significant differences were observed in several categories, including Trustworthiness (H = 31.243, p = .000), Security (H = 15.607, p = .004), Toxicology (H = 10.134, p = .038), and Social Norms (H = 19.037, p = .001). However, no statistically significant difference was found for Bias (H = 3.413, p = .491).

These results indicate that participant perceptions vary significantly across different groups for most categories related to AI ethics and social norms, with the exception of Bias. The significant p-values suggest meaningful variations, particularly in the domains of Trustworthiness and Social Norms, implying that the sampled groups have different views on how AI systems like ChatGPT align with ethical principles and societal expectations. Therefore, contradictory and occasionally neutral opinions demonstrated the necessity of qualitative research as never before. It appeared that a revision of the material from the perspective of experts was necessary. In this study, we accepted as experts those who had worked as AI developers for at least three years after graduating with a Master's degree in computer science or a Ph.D. in a computer science-related discipline from a university professor.

\begin{table*}[h]
\centering
\caption{Test Statistics\textsuperscript{a,b}}
\label{Table3}
\begin{tabular}{lccccc}
\toprule
 & Bias & Trustworthiness & Security & Toxicology & Social Norms \\
\midrule
\textbf{Kruskal-Wallis H} & 3.413 & 31.243 & 15.607 & 10.134 & 19.037 \\
\textbf{df} & 4 & 4 & 4 & 4 & 4 \\
\textbf{Asymp. Sig.} & 0.491 & 0.000 & 0.004 & 0.038 & 0.001 \\
\bottomrule
\end{tabular}
\begin{flushleft}
\textsuperscript{a} Kruskal Wallis Test \\
\textsuperscript{b} Grouping Variable: Ethic
\end{flushleft}
\end{table*}

\subsection{Qualitative Analysis: Interview Study }

\subsubsection{Bias in ChatGPT}
\hfill \break

Our study shows that Bias in the LLM, such as ChatGPT is a complex issue that can be viewed from multiple perspectives, including the everyday user's experience, gender, race, and data. Although most participants confirm ChatGPT has bias, some regular users mentioned ChatGPT may seem largely neutral, especially those who don't venture into controversial or sensitive topics. As P7 noted, \enquote { \emph {I don't see any bias in normal queries or for academic purposes.}} This sentiment was echoed by several other participants (P3, P5, P6), who felt that for common tasks, the AI performed well. In this kind of everyday usage, biases are less noticeable, and users might feel that ChatGPT provides fair and balanced responses.

However, a deeper study of the thematic analysis reveals that the collection and training of data based on specific locations or subjects often embeds biases within these systems. P14 pointed out, \enquote { \emph {ChatGPT is prone to being biased against specific and controversial topics}}, reflecting on how data related to different subjects, genders, and regions isn’t evenly distributed. Four other participants (P4, P5, P19, and P16) also raised concerns about how uneven data distribution can lead to biased results. When the AI addresses certain topics, particularly those not well represented in its training data, this unevenness can lead to skewing of its responses. Like other AI systems, ChatGPT, when primarily trained on data from the U.S. or Western Europe, ends up reflecting societal biases.

Geographic bias is another issue that came up repeatedly. P16 noted, \enquote { \emph {ChatGPT often defaults to U.S.-centric data,}} while P19 added, \enquote { \emph {The answers it gives in Iran or a third-world country are definitely different in quality.}} Altogether, three participants (P16, P19, P6) mentioned how geographic differences affect the relevance and accuracy of ChatGPT's answers. These discrepancies reflect deeper global inequalities, where certain areas or cultural contexts are prioritized, leaving others underrepresented.

Gender bias was also mentioned by a few participants. For example, P16 observed that \enquote { \emph {LGBTQ+ individuals feel misunderstood by ChatGPT when seeking mental health support,}} pointing to broader issues in how AI handles sensitive topics related to gender and identity. Another participant, P14, said that while they hadn’t personally noticed gender bias, \enquote { \emph {it likely exists to some extent.}} Overall, two participants (P16, P14) highlighted concerns related to gender, although it was clear that this bias was more subtle than others.

Participants highlighted that a practical way forward is to make training datasets more transparent and diverse. AI systems like ChatGPT must incorporate data from various regions, languages, and cultures to provide more balanced responses. As P14 suggested, \enquote { \emph {It’s about making sure the data represents everyone, not just a specific group.}} Additionally, P19 emphasized the importance of localizing AI to cater to specific needs, stating, \enquote { \emph {We need AI that considers local cultures and contexts, not just Western viewpoints.}} Expanding the training scope and implementing more stringent checks for fairness in outputs can help address these concerns.

\subsubsection{Trustworthiness}
\hfill \break

Trustworthiness in AI chatbots, such as ChatGPT, was a major topic of discussion among participants, with a variety of views on how much confidence can be placed in these systems. Many participants highlighted that because chatbots are based on probabilistic models, they can’t always be trusted to provide accurate or consistent answers. P35 remarked, \enquote { \emph {AI is based on probabilities, not certainties,}} cautioning against the expectation of perfect answers. Similarly, P12 referred to chatbots as \enquote { \emph {pattern-matching machines,}} which, while useful, are not inherently trustworthy like humans. This concern was shared by three other participants (P19, P33, P36), who agreed that although chatbots generally try their best, users should be aware of their limitations. Despite these concerns, some participants felt that users could still develop a level of trust as long as they understand that occasional mistakes are inevitable.

A significant issue raised by five participants (P24, P30, P33, P36, P9) was the need for transparency and explainability in AI systems to foster greater trust. P36 stressed that \enquote { \emph {for users to trust chatbots, they need clear explanations about how the chatbot operates and how it generates its responses.}} Participants noted the lack of insight into the system's decision-making process, which made them cautious about fully trusting the results. P30 pointed out that \enquote { \emph {while users interact with chatbots regularly, they don’t always understand what’s happening behind the scenes, making it harder to develop full confidence in the system.}} P24 echoed these sentiments, adding that users aren’t aware of the data fed into the system or the mechanisms driving the chatbot’s responses.

Concerns about data transparency also played a role in shaping participants’ trust in chatbots. P16 raised concerns about users not fully understanding how their data is collected, stored, or used by the chatbot, which led to skepticism. Five participants (P7, P9, P16, P19, P30) expressed a desire for more transparency around data usage. P7 suggested that \enquote { \emph {if chatbots were more open about the data they use and the sources behind their responses, it could significantly improve user trust.}} P9 proposed that open-source AI models might be a way forward, as they would allow users to see how the chatbot works and what data it relies on, thereby fostering greater trust.

Reliability was another recurring theme, with many participants pointing to instances where the chatbot provided inconsistent or incorrect information. While some participants, such as P34, acknowledged that chatbots can be helpful and accurate in many cases, five other participants (P33, P34, P16, P30, P19) emphasized that inconsistent responses can erode trust. P34 shared how asking the same question in different ways could lead to conflicting or inaccurate responses. P33 highlighted the importance of users double-checking the chatbot's output, especially when it comes to critical or sensitive tasks, since the tool is not yet reliable enough to be trusted without human oversight.

To address these concerns and increase trustworthiness, participants agreed on several key solutions. First, increasing transparency around how AI systems like ChatGPT operate was identified as essential. Providing users with more detailed information about how the chatbot generates its answers, along with clearer explanations of the data sources it relies on, could help users better understand its limitations. Additionally, offering more insight into how user data is handled and ensuring stringent data privacy measures would further bolster trust, as suggested by P16 and P7. Secondly, continuous evaluation and feedback mechanisms should be in place to refine and improve chatbot performance. P33 emphasized that user feedback should be integrated into system updates to address issues like bias and inaccuracy. Finally, exploring open-source models, as proposed by P9, could allow users to verify how AI systems function, fostering greater transparency and accountability in their development and usage. These steps, combining transparency, reliability, and data privacy, were seen as vital for building a more trustworthy relationship between users and AI systems.

\subsubsection{Security in ChatGPT}
\hfill \break

Participants expressed a range of concerns about the security of their data when using ChatGPT, with several highlighting uncertainties surrounding how personal information is handled. P38 mentioned, \enquote { \emph {I haven’t noticed any security issues with ChatGPT, but since it collects users' data, there might be some concerns. They should definitely improve security and offer options for users to protect their information.}} P34 compared ChatGPT’s security to platforms like Google or Facebook, saying, \enquote { \emph {From my perspective, I don't know because I don't have access to that information. I don't think it's any less secure than using the internet via Google or Facebook.}} These comments reflect the broader unease participants feel about how AI platforms handle sensitive information, emphasizing the need for more transparency (P38, P34).

Several participants drew comparisons between ChatGPT and other AI systems, raising concerns about OpenAI’s partnerships with major tech companies. P32 observed, \enquote { \emph {I find Gemini's security better than that of OpenAI. The story is that OpenAI integrates with various companies[...] which implies to me that it’s overly open since it runs on both iOS and Microsoft.}} P33 added, \enquote { \emph { The web interface does log all data in users’ personal profiles,}} showing a concern that as AI systems become more interconnected, their security may become more vulnerable (P32, P33).

In terms of improving security, some participants suggested specific technical solutions and greater transparency. P8 recommended, \enquote { \emph {AI providers should consider separate servers for each user to maintain the confidentiality of their data.}} P38 echoed the need for transparency, suggesting, \enquote { \emph {If you can prove this to users while still maintaining the same capabilities as ChatGPT, it could help foster greater trust.}} Participants believed that practical solutions and better communication about data handling could improve security and foster trust (P8, P38).

Regulation was also seen as key in ensuring AI security. P22 said, \enquote { \emph {I haven’t delved deeply into how OpenAI is addressing this issue. But I think they have some compliance regulations in their work.}} P19 stressed the need for strong regulatory oversight: \enquote { \emph {This should become a major issue because it’s vital, in my opinion, and it could be dangerous if not handled properly.}} These participants underscored the importance of external regulations and corporate responsibility in ensuring AI platforms adhere to ethical and security standards (P22, P19).

Participants suggested enhancing transparency and implementing stronger security measures as key solutions. AI companies like OpenAI should ensure encryption and other protective mechanisms while also being more open about their data-handling practices. P7 noted, \enquote { \emph {There is encryption behind ChatGPT, so all of your identity information would be stored in an encrypted version, not a plain version.}} By combining technical safeguards like encryption with transparency about data collection, storage, and usage, AI companies could build more trust and confidence among users ( P7, P38).

\subsubsection{ Toxicology and Hate Speech in ChatGPT}
\hfill \break

Participants expressed significant concerns about the potential misuse of AI, particularly tools like ChatGPT. For example, some highlighted how these tools could easily be exploited for harmful activities. P38 mentioned, \enquote { \emph {I haven’t used the open APIs from ChatGPT, but I’ve heard that ChatGPT provides APIs for users to use language models. In that case, such models could potentially be used in phishing, scamming, or gathering unauthorized information from people.}} Another participant, P33, likened the situation to a cat-and-mouse game, saying, \enquote { \emph {You have the black hat and white hat hackers. The black hat hackers are going to try and use AI to access secure information[...] and the white hat hackers are going to use AI to try and prevent it.}} (P33, P38).

Several participants raised concerns about how prompt engineering allows users to manipulate AI systems like ChatGPT. For instance, P32 shared, \enquote { \emph {There have been instances when the guardrail was bypassed by asking the AI to produce creative writing and detail how the main character conducts a malicious activity}}. Similarly, P8 stated that \enquote { \emph {It’s very moral and educated by default, but someone who knows prompt engineering can receive harmful content.}} These responses indicate how skilled users can exploit AI systems for unintended purposes (P8, P32).

Concerns about hate speech and misinformation were common. P12 noted, \enquote { \emph {Hate speech can adapt and evolve very quickly and introduce new terminology[...] so you have to have AI that’s consistently being trained on whatever emerging terminology is being used.}} This sentiment was echoed by P36, who emphasized the challenge of applying real-time detection: \enquote { \emph {While it’s possible, we don’t know how soon or effectively it will happen.}} Participants recognized that while ChatGPT and other AI systems are improving, they struggle to keep up with the rapid evolution of harmful content (P12, P36).

The spread of misinformation through AI was a prominent concern for participants. P5 described how ChatGPT could easily be used to generate fake news, stating, \enquote { \emph {For example, we can give it a sentence like 'the coronavirus spreads in September,' and tell ChatGPT to make ten sentences with different modes based on this. It will do this easily, and with fake accounts, we can spread misinformation in various ways.}} The ease with which AI can generate misleading content highlights the need for stronger safeguards to prevent such harmful uses (P5).

Participants shared a range of thoughtful ideas on how to make AI more ethical and secure. P37 highlighted the importance of constant system updates and user education to help prevent cyberattacks: \enquote { \emph {We need constant monitoring, penetration testing, regular updates, and user education.}} P15 suggested using anomaly detection to identify and respond to suspicious activities, saying, \enquote { \emph {Using advanced anomaly detection algorithms to identify and respond to suspicious activities.}} These responses suggest a need for both technical solutions and strong ethical guidelines to ensure AI is used responsibly (P15, P37).

\subsubsection{Social Norms in ChatGPT}
\hfill \break

During the interviews, participants from Iran, Germany, and the US shared varied perspectives on who should be responsible for ChatGPT’s outputs. Many felt that OpenAI, as the system’s creator, should bear the primary responsibility. A participant from Germany (P3) emphasized, \enquote { \emph {At the end of the day, OpenAI created the system, so they should be accountable for its responses.}} However, others argued that responsibility should also be shared with users. For example, P16 from Germany noted, \enquote { \emph {There’s a balance—OpenAI needs to ensure ethical use, but users should also understand how to responsibly use the AI.}} This reflects a broader sentiment that while developers must set ethical standards, users must also exercise caution and critical thinking when interacting with AI tools.

A recurring concern, particularly among participants from non-Western regions, was the issue of cultural biases embedded in AI. P35 from the US highlighted that when asking an AI system about an image, ChatGPT provided a more Western interpretation, while another AI system leaned toward an Eastern perspective. Similarly, P38 from the US pointed out that users from countries like Iran may receive less culturally relevant responses, given that the AI’s training data is predominantly Western. Moreover, participants from Iran raised additional concerns about limited access to AI due to government restrictions, with P21 noting that \enquote { \emph {Sanctioning countries like Iran [...] for the security of the program. I was able to access it with a different phone number, and certainly, others can do the same with other phone numbers}}. This suggests that in regions with restricted internet access, the lack of reliable AI tools may further disadvantage users, both in terms of accessibility and relevance of content.

When discussing autonomy, participants had mixed views on whether ChatGPT supports or undermines user independence. A participant from Germany (P27) described how ChatGPT made tasks like writing much easier and less time-consuming: \enquote { \emph {It gives me more freedom, especially with writing, where I used to waste hours.}} However, a participant from the US (P19) expressed concern about the risk of over-reliance on AI, particularly for younger generations, stating that it might impair their critical thinking skills: \enquote { \emph {People, especially students, might become too dependent on it.}} Additionally, participants from Iran, such as P20, raised concerns about the potential for governments to misuse open-source AI technology, exacerbating issues like censorship and surveillance. These concerns reflect a tension between the empowerment that AI tools can offer and the potential risks they pose to intellectual autonomy and privacy, particularly in more restrictive environments.

Another important topic was the global versus localized design of AI. P17 from Iran expressed optimism about AI’s ability to bridge cultural divides, hoping for a future where systems like ChatGPT could facilitate global understanding. However, many participants, including P35 from the US, acknowledged that AI is still biased towards Western perspectives, making it less relevant or accurate for users in non-Western countries. This highlights a significant gap in AI’s ability to serve all users equitably, with participants from Iran further emphasizing the challenges they face due to internet censorship, where accessing AI tools requires VPNs, adding yet another layer of difficulty. Participants from Iran voiced concerns that AI systems are not fully equipped to address the specific needs and constraints of users in more controlled or censored regions.

When it came to solutions, participants shared thoughtful recommendations to boost trust and usability in AI systems like ChatGPT. P14 from Iran, for instance, emphasized the importance of transparency, calling for clear communication about the limitations and biases of these tools. P37 from the US suggested regular updates and improved user education on how to responsibly interact with AI to help prevent misuse. Additionally, P15 from Iran proposed more robust technical measures, such as advanced algorithms that could detect and mitigate biases, as well as prevent the misuse of AI by governments. Across the board, participants agreed that the future of AI depends not only on better technology but also on informed and responsible use, especially in regions where access to these tools is limited or controlled by external factors like government intervention.

\subsubsection{ Ethical Data in ChatGPT}
\hfill \break

The study of the ethical aspects of data norms linked to ChatGPT has resulted in a complex perspective from experts in the fields of AI and empirical research. An important subject that arose was the conflict between concerns about security and the usefulness of AI systems. A few experts expressed concerns over ChatGPT's methods of collecting data. As one expert stated: \enquote { \emph {Really, if you look at it, it is not moral at all[...] It is not ethical, it means that someone who comes puts some data on the internet... Most likely, he does not like his data to be collected by other people}} (P6). This perspective highlights concerns over privacy and the ethical implications of data collection. Another expert emphasized, \enquote { \emph {if you look at it, it is not moral at all. At all, collecting data at the network level}} (P5), underscoring the ethical dilemma of collecting user data without explicit consent.

On the other hand, other individuals highlighted the significance of user accountability and motivations, positing that ethical considerations rest with the decisions and behaviors of users while engaging with the technology. As one participant noted: \enquote { \emph {The results you get on the ChatGPT[...] It depends on the user and the ethics they use it}} (P11), indicating that the ethical use of ChatGPT is contingent on how users engage with the system. Furthermore, there was a reoccurring discussion concerning the boundary between legal and moral ethics, emphasizing the complexities of evaluating ethics in the context of AI. When ChatGPT supplied material without appropriate acknowledgment or reference, particularly when drawing from private sources, ethical concerns were raised. One participant pointed out: \enquote { \emph {for every answer, for every information that ChatGPT provided for the user, the user needs to [...] check the accessibility of that reference}} (P7), emphasizing the need for ChatGPT to properly cite its sources. Some participants pointed out that future updates of ChatGPT may address these ethical concerns and improve its standing by enhancing data collection practices. As one expert optimistically noted \enquote { \emph {he takes ethics into account now, but I think that may not be the case in the future}} (P1). Overall, our findings highlight the complex ethical landscape surrounding AI, highlighting the importance of ongoing research and collaboration among AI developers, ethicists, and policymakers to navigate these complexities and ensure that AI systems like ChatGPT follow ethical principles while providing valuable services to users.

\section{Discussion}

\subsection{Ethical Challenges of Large Language Models }

In this article, we discuss several ways to use ChatGPT ethically, including bias, trustworthiness, security, toxicology, social norms, and ethical data. The results of this study shed important light on how users of ChatGPT, an AI-based conversational system, perceive social norms and ethical issues related to its use. Participants used a wide range of LLMs, including Claude, Gemini, Copilot, Leonardo, and Magnific for images; Luma and Runway for video; and a few others like 3D AI Studio. As P24 mentioned, different LLMs and different versions of ChatGPT do not conflict too much. On the other hand, some participants noted differences between ChatGPT and other models. While our current study focused on ChatGPT due to its predominant use among our survey participants, based on this information from our interview with experts, we believe that the ethical considerations outlined are universally applicable to other LLMs. The categories we identified are common concerns in the AI ethics literature across various platforms.

In addition, a thorough knowledge of the ethical issues related to ChatGPT is provided by the combination of a quantitative study and a qualitative examination of an empirical study. Significant discrepancies in the participants' views of ethics for six ChatGPT features were found using the Kruskal-Wallis Test. According to these results, data safety, user privacy, trustworthiness, and social norms all play significant roles in determining how people perceive ethical behavior \cite{Leslie2019Understanding}. The system's biases raised issues among the participants, notably those that relate to local vs. global biases and biases in favor of conservative viewpoints. This emphasizes how crucial it is to deal with biases throughout the design and training stages of AI systems to assure fairness and reduce possible harm \cite{Varona2022Discrimination}. Another study conducted by Das et al. confirms that the ChatGPT model demonstrates inferior performance for the counter-speech-related functionalities, struggling to distinguish between hate speech and counter-speech. In most cases, ChatGPT acts conservatively and responds, "I am sorry, but I cannot determine...” \cite{das2023evaluating}.

Another significant concern raised by participants centers on user safety, as many worry that malicious actors may exploit ChatGPT’s vulnerabilities to target individuals or groups. The use of AI systems responsibly requires robust security measures and stringent ethical guidelines to address these concerns \cite{Ryan2020Artificial}. The qualitative research gave further insights into users' perceptions of ChatGPT's ethical standards and social norms. Participants stressed the value of accurate data, understandability, and the human element in decision-making. They stated a need for AI systems to accurately present facts and offer concise justifications for their results. This demonstrates the requirement for accountability and openness in AI systems so that people can comprehend and trust the system's decision-making process.

In addition, the participants emphasized the centrality of human oversight, ethical standards, and expert engagement to guide AI systems such as ChatGPT. These findings align with the responsible AI development framework proposed by Dignum \cite{Dignum2017Responsible}, emphasizing the embedding of human values, transparent processes, and social norms into AI design. By adhering to established ethical principles—openness, user safety, sound decision-making, and interpretability—AI developers can better align technological capabilities with societal values. Overall, our study contributes to ongoing debates on AI ethics by offering both quantitative and qualitative evidence that underscores the importance of mitigating biases, strengthening user safety, improving data integrity, and maintaining the human element in AI-mediated interactions. Embracing these principles and further investigating how they can be effectively operationalized will help ensure AI systems like ChatGPT continue to evolve in ethically sound and socially beneficial ways.

\subsection{Principles of AI Ethics and Social Norms}

ChatGPT can perpetuate and amplify societal biases if it is not designed and tested carefully \cite{hermann2022leveraging}. Indeed, chatbots have been shown to exacerbate race-specific social harm. Some researchers suggest that focusing on race-sensitive innovations in chatbots and AI could help mitigate these biases \cite{schlesinger2018let}. Our quantitative findings indicate a wide range of user perceptions regarding bias. Meanwhile, our qualitative insights reveal three distinct perspectives. The first perspective, primarily from Iranian participants, attributes bias to the nature of the training data, which is predominantly collected from Western sources. Participants described how collecting data from one region can lead to biased responses and disregard certain social norms, ultimately eroding user trust in LLMs. The second perspective, drawn mainly from US and German participants, highlights that bias stems from broader societal biases, such as race and gender. Furthermore, experts across countries share a common view on the importance of data quality, advocating for precise, context-relevant training data to ensure accurate and unbiased outcomes. The ChatGPT initiative attempts to account for contextual factors and adapt to the user's geographical parameters. Consequently, LLMs employing reinforcement learning with human feedback—an extra training layer that involves human input—may partly address these challenges \cite{Abdullah2022ChatGPT}.

Our research shows that transparency and user perception of a trustworthy agent like ChatGPT impact society's consideration of an LLM as an ethical agent or not. Regarding LLM as a black box, our result shows that companies should be transparent regarding the data they collect, how it is utilized, and the GPT-based algorithms and decision-making processes \cite{rivas2023marketing}. As previous researchers have shown, transparency and explainability enhance the trustworthiness of LLMs \cite{10.1145/3584931.3607492, barman2024beyond, poslonoutpacing, quttainah2024cost}. According to quantitative data, a significant proportion of participants expressed ambiguity or disagreement regarding ChatGPT's ethical behavior, contextual comprehension, and potential hazards. Also, social norms can impact how users understand LLM output. In the research, most of the participants from Iran (P5, P28) mentioned our understanding can be different about output; this is because we use different languages and should be considered to develop LLM. On the other hand, most researchers define consistency as models that may generate different outputs even with the same input \cite{lee2024one}. Based on our findings about how social norms should impact the output of LLMs for different users with different cultures, this definition should be redefined. Qualitative insights highlight the complexities of AI systems, highlighting the significance of data quality, transparency, and interpretability in determining reliability. Some participants advocated for open-source models to increase transparency, citing privacy and security concerns regarding data sharing and inaccurate information production.

In addition, a substantial majority of participants favored ethical constraints on AI’s content generation, prioritizing human interests over nonhuman ones, yet expressed varying levels of confidence in ChatGPT’s data accuracy. In our qualitative expert interviews, participants stressed the need for explicit privacy boundaries during AI interactions, ethical content filtering, and user permissions, supported by continuous user feedback \cite{derner2023beyond}. Hate speech is defined as any kind of abuse directed against a protected group or its members just because they belong to that group. In accordance with the worldwide legal agreement, protected categories include those based on age, disability, gender identity, race, national or ethnic origin, religion, sex, or sexual orientation \cite{das2023evaluating}. As noted in previous research \cite{das2023evaluating, das-etal-2024-evaluating}, participants in this study highlighted the evolving nature of AI responses to hate speech, reinforcing the need for ethical standards, model updates, and ongoing vigilance. Moreover, our interviews indicate that participants from Iran often link toxicological issues to manipulative prompt engineering, whereas participants from the US and Germany primarily emphasize misinformation and malicious behavior. Despite these differences, cyber threats remain a common concern across all regions, and experts propose guidelines for both users and developers, continuous system updates to detect emerging threats, and clear boundaries for LLM access. 
 
Our mixed-methods data on ChatGPT’s security reveals a diverse array of viewpoints. Quantitatively, most participants voiced worries about security vulnerabilities, potential misuse, and whether the system effectively safeguards user data—concerns consistent with research indicating new NLP paradigms introduce security threats \cite{shi2023badgpt, 10.1145/3706599.3720152}. Qualitatively, participants expressed apprehension about AI’s expanding influence, advocating for stricter regulations and transparency to address present security weaknesses and the need for ongoing evaluations. Although OpenAI’s policy states that non-API data is employed for improving services while removing personally identifiable information, the security of sensitive data remains a concern, prompting users to avoid sharing private or exploitable information with ChatGPT \cite{derner2023beyond}. Still, some participants remain optimistic, pointing to a lack of reported major security breaches and the use of encryption in data protection. Suggestions from participants included separating user servers and boosting user awareness. Overall, these findings highlight the importance of tackling security risks while acknowledging existing safeguards and incorporating user engagement in data protection. Continued assessment and enhancement of ChatGPT’s security protocols remain crucial.

In our broader examination of AI ethics and social norms, many Iranian participants acknowledge that LLMs—especially ChatGPT-provide significant benefits, such as improving productivity and quality of life. However, all participants in the US and Germany noted that LLMs can also have adverse effects, particularly regarding innovation and legal issues, suggesting a need to revise copyright and anti-plagiarism policies. Without such revisions, as other researchers mentioned, crises may emerge in creative fields like music, painting, photography, and video production \cite{samuelson2023generative, sag2023copyright}. In addition, LLMs can influence different national economies, with potential impacts on livelihoods, as noted by participants from Iran, ultimately reshaping social norms and entire communities. Consequently, our participants underscored the importance of human oversight and accountability. Experts highlighted the pivotal responsibility of AI developers (particularly OpenAI) in taking responsibility for AI outputs and ensuring active monitoring. Views on ChatGPT’s impact on personal autonomy vary, with some participants observing minimal encroachment, while others fear growing dependence and potential social manipulation. Regulatory measures and educational initiatives were suggested to manage these risks, even as some experts viewed AI chatbots as tools for augmenting rather than diminishing autonomy. Given that ChatGPT has yet to achieve artificial general intelligence (AGI), despite its advanced writing capabilities and sophisticated model, many fundamental aspects of AGI—such as self-awareness, emotion, motivation, metacognition, and deliberative processes—remain absent \cite{goertzel2014artificial}. A number of Iranian participants specifically noted that religious and cultural ideologies should influence LLM outputs; otherwise, globalized Western ideologies might overshadow smaller cultures, potentially causing them to disappear over time. Overall, discussions of AI ethics and social norms underscore that ethical principles must be contextualized locally and that training data quality should adapt accordingly.

\begin{figure*}[h]
  \centering
  \includegraphics[width= \linewidth]{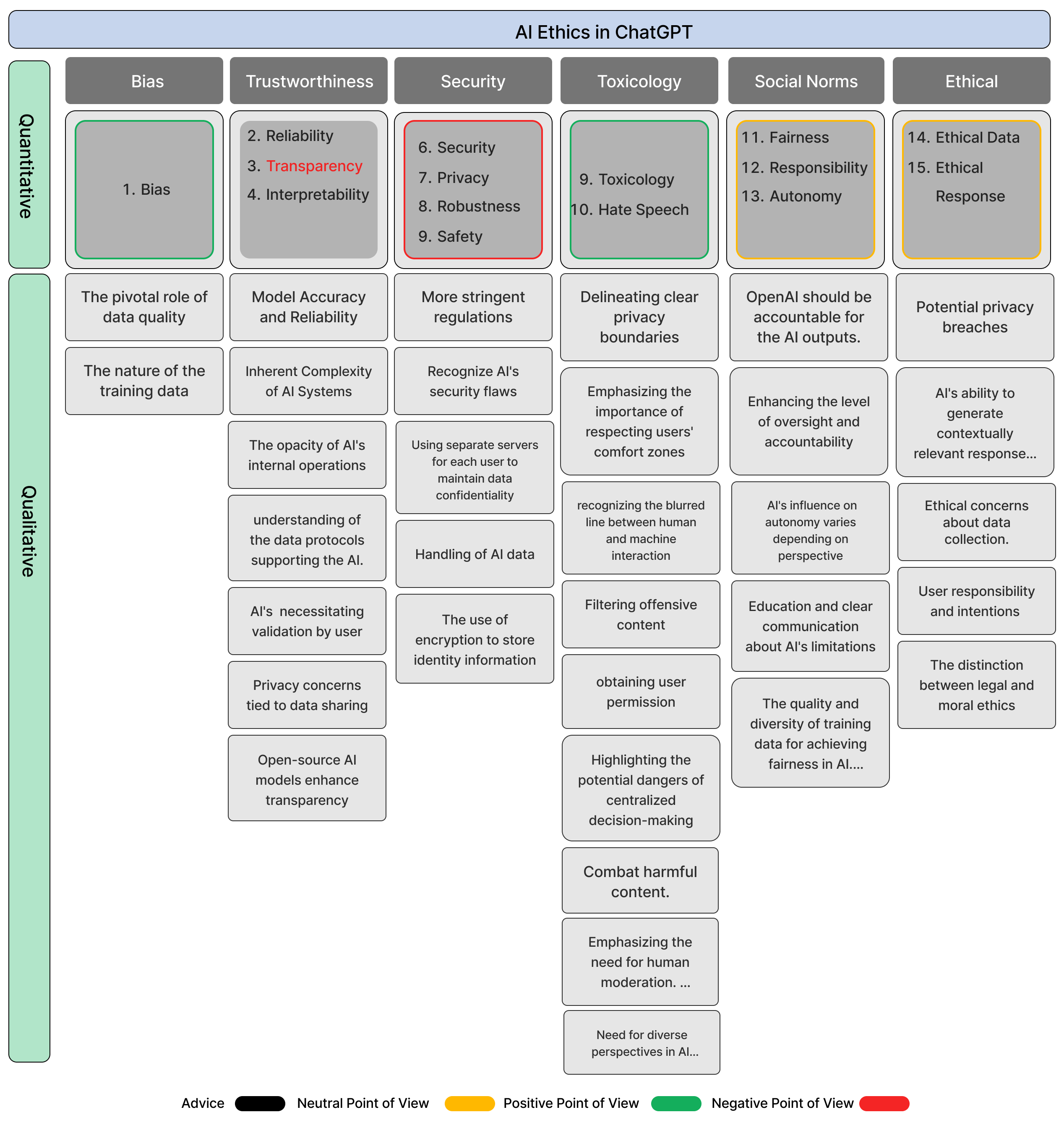}
  \caption{Highlighted points of the ethical framework}
    \label{Fig3}
\end{figure*}

Ethics and AI have been the subject of ongoing discussion, with many globally convergent concepts including transparency, justice, and fairness \cite{jobin2019global}. Quantitatively, our findings underscore the diversity of opinions regarding the morality of ChatGPT's data collection practices, with a significant proportion expressing neutrality. Furthermore, the data indicates a prevailing uncertainty surrounding the ethical application of ChatGPT's results, shedding light on the complexity of public perceptions regarding AI ethics. Unfortunately, the development and implementation of AI ethics have been unsuccessful and are frequently viewed as an arbitrary, non-binding framework imposed by institutions outside the tech sector \cite{hagendorff2020ethics}. \autoref{Fig3} demonstrates our framework for evaluating the ethical development of AI systems and how we can develop software such as ChatGPT, etc. Some experts voiced apprehensions about potential privacy breaches arising from the system's ability to generate contextually relevant responses based on user chat history, raising ethical concerns about data collection. Concurrently, others emphasized user responsibility and intentions, positing that ethicality may hinge on user choices and behaviors during interactions with the tool. Furthermore, the distinction between legal and moral ethics surfaced as a recurring point of discussion, underscoring the intricate nature of ethical assessments in the AI domain. Such efforts are vital to ensuring that AI systems like ChatGPT align with ethical principles while continuing to provide valuable services to users."

To summarize, bias in LLMs stems from both dataset limits and larger cultural preconceptions, emphasizing the need of rigorous data quality metrics and reinforcement learning with human input. Transparency, explainability, and open-source techniques may build user confidence, yet cultural and language issues challenge the idea of consistency and need flexible AI solutions. Strong ethical limits and content regulation are critical to protecting users, with cyber dangers acknowledged as a global problem necessitating ongoing monitoring and system improvements. Although ChatGPT and related LLMs provide significant advantages (e.g., increased productivity), they also present issues in innovation, legal frameworks (e.g., copyright), and cultural preservation, emphasizing the requirement of human supervision, responsibility, and context-specific training data. Finally, bridging the gap between theoretical AI ethics principles and actual enforcement is critical, given that moral and legal viewpoints often overlap, and user intentions have a substantial impact on ethical consequences.

\subsection{Implications for the CSCW Community}

Many studies in the CSCW communities studied about how security impacts the CSCW community in the AI era \cite{yi2003privacy, 10.1145/3500868.3559706, 10.1145/3462204.3481724, 10.1145/2998181.2998191, 10.1145/2675133.2675225}. But except a few papers, you cannot find an investigation about how LLMs impact security in the CSCW community \cite{yi2003privacy}. In our study, we realized users and experts are mostly concerned about data policies and policies from companies related to data collection. Users believe that OpenAI should revise its policy for data collection, as they perceive LLMs as a black box and find it difficult to understand how LLMs interact with data. These policies impact the many cooperatives that used AI as an assistant. As the CSCW researchers mentioned, researchers face obstacles regarding the ethical implications of collecting and analyzing publicly accessible yet potentially sensitive online data \cite{vitak2016beyond}. Additionally, we discovered a privacy and safety issue with LLMs during our study, which could potentially impact the CSCW communities involved in cooperative work. Based on the research from the CSCW, the agent-supported collaborative work paradigm has the ability to facilitate information sharing among participants, provide assistance to users, enhance processes and procedures, and deliver user-friendly interfaces for CSCW systems. As computer programs, agents often lack the cognitive capacity to recognize harmful behaviors comparable to human intelligence. Consequently, the security of agent-supported collaborative work like ChatGPT has significant importance \cite{yi2003privacy}. Most experts mentioned human supervision and changing policy as effective solutions for security problems.

In the CSCW communities, many researchers focus on the different types of bias, including cognitive bias \cite{10.1145/3584931.3611284}, data bias \cite{10.1145/3462204.3481729} and so on. Attention to the bias is important because, based on our research, it effects trustworthiness and social norms. On the other hand, if data bias happened, the user lost trust in the system, and this is important for cooperative works and human AI collaboration \cite{10.1145/3584931.3611284}. In our study, different types of bias were also mentioned by our participants, including gender, race, data, location, and language. For instance, Iranian users reported experiencing location bias and access limitations from OpenAI, which prompted them to use virtual private networks to change their IP addresses. Conversely, American and German users primarily brought up issues related to gender and race. For example, one expert mentioned ChatGPT responds to the sensitive question of health care just by adding information about LGBTQ. Also, in our study about trustworthiness and toxicology, we faced with the different problems that can impact the CSCW community. Most and foremost, users perceived that due to consistency in LLMs, some hallucinations happened, and as a result, this impacted their result due to the inconsistent output of ChatGPT \cite{10.1145/3584931.3608438}. Users mentioned they need the system to be transparent in the data, user, and output. And being a black box LLM without a clear policy from the company impacts their trust, and they suggested preparing open source LLMs and more human feedback loops can increase trustworthy to work with the AI agent in everyday work \cite{10.1145/3500868.3559450}. On the other hand, some of the Iranian participants are concerned about the abuse of open-source LLMs by the government for hacking and attacking other countries.

\subsection{Limitation and Future Work}

The evaluation of ChatGPT's ethics and social norms is dependent on the model version and content restrictions in effect at the time of the study. Our participant sample is limited to long-term users rather than first-time users and inactive users. Although the long-term user sample provides the study with detailed insights based on their extensive experience and familiarity with ChatGPT, first-time users may offer different perspectives due to their fresh expereince, and inactive users may provide valuable reasons for discontinuing use. These varied responses could yield different but interesting results, warranting further investigation in future research.

As technology and safeguards evolve, so may the risks and vulnerabilities. Second, the paper concentrates predominantly on ChatGPT and the security concerns associated with it. While ChatGPT serves as a representative example, it may not incorporate the full range of LLMs and their unique risks. Future research could examine a broader spectrum of LLMs in order to provide a deeper understanding of the ethics and social norms. Finally, it is critical to continue investigating the human aspect in AI interactions and decision-making. Future research could delve deeper into the role of human governance, ethical standards, and domain experts in shaping the behaviour and decisions of AI systems. This would facilitate the development of AI systems compatible with human values and ethical concerns. This study provides the framework for comprehending ChatGPT's social norms and ethical consequences.

\section{Conclusion and Future Work}

In this study, we investigated how users of ChatGPT, an AI-based conversational system, perceived social norms and ethical issues related to its use. We were able to learn more about the ethical problems with ChatGPT by using both quantitative research (using the Kruskal-Wallis Test) and qualitative analysis of interview data. The results shed light on crucial decision-making factors such as bias, trustworthiness, security, toxicology, hate speech, social norms, and ethical data that influence users' ethical judgments. Our quantitative research revealed statistically significant differences in the participants' ethical assessments of several ChatGPT features. These findings demonstrate the need to consider trustworthiness, user security, and data veracity throughout the design and implementation of AI systems. In addition, the qualitative study provided in-depth, complex perspectives from AI experts, which helped us better understand the ethical implications and issues surrounding the use of ChatGPT. We have provided a mixed-method-based guideline for social norms and ethics in LLM systems. These guidelines emphasize the importance of providing openness, user safety, responsible decision-making, and understandability as a top priority when developing and employing AI systems such as ChatGPT. By incorporating these concepts into ethical frameworks, we can promote trust, accountability, and ethical behavior in the field of AI.

Our qualitative data revealed that a substantial proportion of participants expressed concerns about bias in ChatGPT, noting uneven performance tied to geographic and cultural differences. Many also indicated uncertainty regarding trustworthiness, particularly in relation to data transparency and potential inaccuracies. Security emerged as another focal point, with a significant number of respondents voicing worries about data handling and unauthorized access. Notably, the semi-structured interviews highlighted a broad awareness of ChatGPT’s potential for generating hate speech to facilitate misinformation, underscoring the need for ongoing oversight and ethical guidelines. The quantitative findings, combined with our qualitative insights, contribute a more nuanced perspective on user experiences with LLMs. By pinpointing specific challenges, such as bias in non-Western contexts, limited data transparency, and security risks, we build upon prior studies’ broader discussions of AI ethics and provide tangible evidence of where user-centered improvements are most urgently required. 

In future research, researchers can extend our analysis to include a broader array of LLMs, such as Google Bard/Gemini, Claude, Meta Llama, and others, to compare their relative biases and behaviors. Future studies should examine whether user perceptions of ethical concerns alter as AI systems evolve with various versions of LLMs. Although our results did not reveal any substantial differences between ChatGPT-3.5 and ChatGPT-4 users, the potential for a shift in perspectives may result from ongoing enhancements to model capabilities. A comparative study that examines the evolution of trust, bias, and ethical considerations in AI models with a concentration on various LLM versions over time, potentially across multiple platforms, would offer valuable insights.

\bibliographystyle{ACM-Reference-Format}
\bibliography{sample-base}


\begin{thebibliography}{110}

%%% the Web address.

\ifx \showCODEN    \undefined \def \showCODEN     #1{\unskip}     \fi
\ifx \showDOI      \undefined \def \showDOI       #1{#1}\fi
\ifx \showISBNx    \undefined \def \showISBNx     #1{\unskip}     \fi
\ifx \showISBNxiii \undefined \def \showISBNxiii  #1{\unskip}     \fi
\ifx \showISSN     \undefined \def \showISSN      #1{\unskip}     \fi
\ifx \showLCCN     \undefined \def \showLCCN      #1{\unskip}     \fi
\ifx \shownote     \undefined \def \shownote      #1{#1}          \fi
\ifx \showarticletitle \undefined \def \showarticletitle #1{#1}   \fi
\ifx \showURL      \undefined \def \showURL       {\relax}        \fi
% invisible to TeX
\providecommand\bibfield[2]{#2}
\providecommand\bibinfo[2]{#2}
\providecommand\natexlab[1]{#1}
\providecommand\showeprint[2][]{arXiv:#2}

\bibitem[Abdullah et~al\mbox{.}(2022)]%
        {Abdullah2022ChatGPT}
\bibfield{author}{\bibinfo{person}{Malak Abdullah}, \bibinfo{person}{Alia Madain}, {and} \bibinfo{person}{Yaser Jararweh}.} \bibinfo{year}{2022}\natexlab{}.
\newblock \showarticletitle{ChatGPT: Fundamentals, applications and social impacts}. In \bibinfo{booktitle}{\emph{2022 Ninth International Conference on Social Networks Analysis, Management and Security (SNAMS)}}. IEEE, \bibinfo{pages}{1--8}.
\newblock
\urldef\tempurl%
\url{https://doi.org/10.1109/SNAMS58071.2022.10062688}
\showDOI{\tempurl}


\bibitem[Akbar et~al\mbox{.}(2023)]%
        {akbar2023ethical}
\bibfield{author}{\bibinfo{person}{Muhammad~Azeem Akbar}, \bibinfo{person}{Arif~Ali Khan}, {and} \bibinfo{person}{Peng Liang}.} \bibinfo{year}{2023}\natexlab{}.
\newblock \showarticletitle{Ethical Aspects of ChatGPT in Software Engineering Research}.
\newblock \bibinfo{journal}{\emph{arXiv preprint arXiv:2306.07557}} (\bibinfo{year}{2023}).
\newblock
\urldef\tempurl%
\url{https://doi.org/10.48550/arXiv.2306.07557}
\showDOI{\tempurl}


\bibitem[Akter et~al\mbox{.}(2022)]%
        {10.1145/3500868.3559706}
\bibfield{author}{\bibinfo{person}{Mamtaj Akter}, \bibinfo{person}{Leena Alghamdi}, \bibinfo{person}{Dylan Gillespie}, \bibinfo{person}{Nazmus~Sakib Miazi}, \bibinfo{person}{Jess Kropczynski}, \bibinfo{person}{Heather Lipford}, {and} \bibinfo{person}{Pamela~J. Wisniewski}.} \bibinfo{year}{2022}\natexlab{}.
\newblock \showarticletitle{CO-oPS: A Mobile App for Community Oversight of Privacy and Security}. In \bibinfo{booktitle}{\emph{Companion Publication of the 2022 Conference on Computer Supported Cooperative Work and Social Computing}} (Virtual Event, Taiwan) \emph{(\bibinfo{series}{CSCW'22 Companion})}. \bibinfo{publisher}{Association for Computing Machinery}, \bibinfo{address}{New York, NY, USA}, \bibinfo{pages}{179–183}.
\newblock
\showISBNx{9781450391900}
\urldef\tempurl%
\url{https://doi.org/10.1145/3500868.3559706}
\showDOI{\tempurl}


\bibitem[Aoyagui et~al\mbox{.}(2024)]%
        {aoyagui2024exploring}
\bibfield{author}{\bibinfo{person}{Paula~Akemi Aoyagui}, \bibinfo{person}{Sharon Ferguson}, {and} \bibinfo{person}{Anastasia Kuzminykh}.} \bibinfo{year}{2024}\natexlab{}.
\newblock \showarticletitle{Exploring Subjectivity for more Human-Centric Assessment of Social Biases in Large Language Models}.
\newblock \bibinfo{journal}{\emph{arXiv preprint arXiv:2405.11048}} (\bibinfo{year}{2024}).
\newblock


\bibitem[Barman et~al\mbox{.}(2024)]%
        {barman2024beyond}
\bibfield{author}{\bibinfo{person}{Kristian~Gonz{\'a}lez Barman}, \bibinfo{person}{Nathan Wood}, {and} \bibinfo{person}{Pawel Pawlowski}.} \bibinfo{year}{2024}\natexlab{}.
\newblock \showarticletitle{Beyond transparency and explainability: on the need for adequate and contextualized user guidelines for LLM use}.
\newblock \bibinfo{journal}{\emph{Ethics and Information Technology}} \bibinfo{volume}{26}, \bibinfo{number}{3} (\bibinfo{year}{2024}), \bibinfo{pages}{47}.
\newblock


\bibitem[Barta and Andalibi(2021)]%
        {barta2021constructing}
\bibfield{author}{\bibinfo{person}{Kristen Barta} {and} \bibinfo{person}{Nazanin Andalibi}.} \bibinfo{year}{2021}\natexlab{}.
\newblock \showarticletitle{Constructing Authenticity on TikTok: Social Norms and Social Support on the" Fun" Platform}.
\newblock \bibinfo{journal}{\emph{Proceedings of the ACM on Human-Computer Interaction}} \bibinfo{volume}{5}, \bibinfo{number}{CSCW2} (\bibinfo{year}{2021}), \bibinfo{pages}{1--29}.
\newblock
\urldef\tempurl%
\url{https://doi.org/10.1145/3479574}
\showDOI{\tempurl}


\bibitem[Bauer(2020)]%
        {bauer2020virtuous}
\bibfield{author}{\bibinfo{person}{William~A Bauer}.} \bibinfo{year}{2020}\natexlab{}.
\newblock \showarticletitle{Virtuous vs. utilitarian artificial moral agents}.
\newblock \bibinfo{journal}{\emph{AI \& SOCIETY}} \bibinfo{volume}{35}, \bibinfo{number}{1} (\bibinfo{year}{2020}), \bibinfo{pages}{263--271}.
\newblock


\bibitem[Bennett(2019)]%
        {bennett2019investigating}
\bibfield{author}{\bibinfo{person}{Sarah~Joy Bennett}.} \bibinfo{year}{2019}\natexlab{}.
\newblock \showarticletitle{Investigating the role of moral decision-making in emerging artificial intelligence technologies}. In \bibinfo{booktitle}{\emph{Companion Publication of the 2019 Conference on Computer Supported Cooperative Work and Social Computing}}. \bibinfo{pages}{28--32}.
\newblock


\bibitem[Boonprakong et~al\mbox{.}(2023)]%
        {10.1145/3584931.3611284}
\bibfield{author}{\bibinfo{person}{Nattapat Boonprakong}, \bibinfo{person}{Gaole He}, \bibinfo{person}{Ujwal Gadiraju}, \bibinfo{person}{Niels van Berkel}, \bibinfo{person}{Danding Wang}, \bibinfo{person}{Si Chen}, \bibinfo{person}{Jiqun Liu}, \bibinfo{person}{Benjamin Tag}, \bibinfo{person}{Jorge Goncalves}, {and} \bibinfo{person}{Tilman Dingler}.} \bibinfo{year}{2023}\natexlab{}.
\newblock \showarticletitle{Workshop on Understanding and Mitigating Cognitive Biases in Human-AI Collaboration}. In \bibinfo{booktitle}{\emph{Companion Publication of the 2023 Conference on Computer Supported Cooperative Work and Social Computing}} (Minneapolis, MN, USA) \emph{(\bibinfo{series}{CSCW '23 Companion})}. \bibinfo{publisher}{Association for Computing Machinery}, \bibinfo{address}{New York, NY, USA}, \bibinfo{pages}{512–517}.
\newblock
\showISBNx{9798400701290}
\urldef\tempurl%
\url{https://doi.org/10.1145/3584931.3611284}
\showDOI{\tempurl}


\bibitem[Bozic and Wotawa(2018)]%
        {Bozic2018Security}
\bibfield{author}{\bibinfo{person}{Josip Bozic} {and} \bibinfo{person}{Franz Wotawa}.} \bibinfo{year}{2018}\natexlab{}.
\newblock \showarticletitle{Security testing for chatbots}. In \bibinfo{booktitle}{\emph{IFIP International Conference on Testing Software and Systems}}. Springer, \bibinfo{pages}{33--38}.
\newblock
\urldef\tempurl%
\url{https://doi.org/10.1007/978-3-319-99927-2_3}
\showDOI{\tempurl}


\bibitem[Briggle and Mitcham(2012)]%
        {Briggle2012Ethics}
\bibfield{author}{\bibinfo{person}{Adam Briggle} {and} \bibinfo{person}{Carl Mitcham}.} \bibinfo{year}{2012}\natexlab{}.
\newblock \bibinfo{booktitle}{\emph{Ethics and science: An introduction}}.
\newblock \bibinfo{publisher}{Cambridge University Press}.
\newblock
\urldef\tempurl%
\url{https://doi.org/10.1017/CBO9781139034111}
\showDOI{\tempurl}


\bibitem[Broersen et~al\mbox{.}(2001)]%
        {Broersen2001BOID}
\bibfield{author}{\bibinfo{person}{Jan Broersen}, \bibinfo{person}{Mehdi Dastani}, \bibinfo{person}{Joris Hulstijn}, \bibinfo{person}{Zisheng Huang}, {and} \bibinfo{person}{Leendert van~der Torre}.} \bibinfo{year}{2001}\natexlab{}.
\newblock \showarticletitle{The BOID architecture: conflicts between beliefs, obligations, intentions and desires}. In \bibinfo{booktitle}{\emph{Proceedings of the fifth international conference on Autonomous agents}}. \bibinfo{pages}{9--16}.
\newblock
\urldef\tempurl%
\url{https://doi.org/10.1145/375735.375766}
\showDOI{\tempurl}


\bibitem[Brusseau(2022)]%
        {Brusseau2022Acceleration}
\bibfield{author}{\bibinfo{person}{James Brusseau}.} \bibinfo{year}{2022}\natexlab{}.
\newblock \showarticletitle{Acceleration AI Ethics, the Debate between Innovation and Safety, and Stability AI's Diffusion versus OpenAI's Dall-E}.
\newblock \bibinfo{journal}{\emph{arXiv preprint arXiv:2212.01834}} (\bibinfo{year}{2022}).
\newblock
\urldef\tempurl%
\url{https://doi.org/10.2139/ssrn.4293514}
\showDOI{\tempurl}


\bibitem[Cecchini et~al\mbox{.}(2024)]%
        {cecchini2024holistic}
\bibfield{author}{\bibinfo{person}{David Cecchini}, \bibinfo{person}{Arshaan Nazir}, \bibinfo{person}{Kalyan Chakravarthy}, {and} \bibinfo{person}{Veysel Kocaman}.} \bibinfo{year}{2024}\natexlab{}.
\newblock \showarticletitle{Holistic evaluation of large language models: Assessing robustness, accuracy, and toxicity for real-world applications}. In \bibinfo{booktitle}{\emph{Proceedings of the 4th Workshop on Trustworthy Natural Language Processing (TrustNLP 2024)}}. \bibinfo{pages}{109--117}.
\newblock


\bibitem[Chandel et~al\mbox{.}(2019)]%
        {Chandel2019Chatbot}
\bibfield{author}{\bibinfo{person}{Sonali Chandel}, \bibinfo{person}{Yuan Yuying}, \bibinfo{person}{Gu Yujie}, \bibinfo{person}{Abdul Razaque}, {and} \bibinfo{person}{Geng Yang}.} \bibinfo{year}{2019}\natexlab{}.
\newblock \showarticletitle{Chatbot: efficient and utility-based platform}. In \bibinfo{booktitle}{\emph{Intelligent Computing: Proceedings of the 2018 Computing Conference, Volume 1}}. Springer, \bibinfo{pages}{109--122}.
\newblock
\urldef\tempurl%
\url{https://doi.org/10.1007/978-3-030-01174-1_9}
\showDOI{\tempurl}


\bibitem[Cristina(2006)]%
        {cristina2006grammar}
\bibfield{author}{\bibinfo{person}{Bicchieri Cristina}.} \bibinfo{year}{2006}\natexlab{}.
\newblock \showarticletitle{The Grammar of Society: The Nature and Dynamics of Social Norms}.
\newblock  (\bibinfo{year}{2006}).
\newblock
\urldef\tempurl%
\url{https://doi.org/10.1017/cbo9780511616037}
\showDOI{\tempurl}


\bibitem[Das et~al\mbox{.}(2023)]%
        {das2023evaluating}
\bibfield{author}{\bibinfo{person}{Mithun Das}, \bibinfo{person}{Saurabh~Kumar Pandey}, {and} \bibinfo{person}{Animesh Mukherjee}.} \bibinfo{year}{2023}\natexlab{}.
\newblock \showarticletitle{Evaluating ChatGPT's performance for multilingual and emoji-based hate speech detection}.
\newblock \bibinfo{journal}{\emph{arXiv preprint arXiv:2305.13276}} (\bibinfo{year}{2023}).
\newblock


\bibitem[Das et~al\mbox{.}(2024)]%
        {das-etal-2024-evaluating}
\bibfield{author}{\bibinfo{person}{Mithun Das}, \bibinfo{person}{Saurabh~Kumar Pandey}, {and} \bibinfo{person}{Animesh Mukherjee}.} \bibinfo{year}{2024}\natexlab{}.
\newblock \showarticletitle{Evaluating {C}hat{GPT} against Functionality Tests for Hate Speech Detection}. In \bibinfo{booktitle}{\emph{Proceedings of the 2024 Joint International Conference on Computational Linguistics, Language Resources and Evaluation (LREC-COLING 2024)}}, \bibfield{editor}{\bibinfo{person}{Nicoletta Calzolari}, \bibinfo{person}{Min-Yen Kan}, \bibinfo{person}{Veronique Hoste}, \bibinfo{person}{Alessandro Lenci}, \bibinfo{person}{Sakriani Sakti}, {and} \bibinfo{person}{Nianwen Xue}} (Eds.). \bibinfo{publisher}{ELRA and ICCL}, \bibinfo{address}{Torino, Italia}, \bibinfo{pages}{6370--6380}.
\newblock
\urldef\tempurl%
\url{https://aclanthology.org/2024.lrec-main.564}
\showURL{%
\tempurl}


\bibitem[Das et~al\mbox{.}(2015)]%
        {10.1145/2675133.2675225}
\bibfield{author}{\bibinfo{person}{Sauvik Das}, \bibinfo{person}{Adam~D.I. Kramer}, \bibinfo{person}{Laura~A. Dabbish}, {and} \bibinfo{person}{Jason~I. Hong}.} \bibinfo{year}{2015}\natexlab{}.
\newblock \showarticletitle{The Role of Social Influence in Security Feature Adoption}. In \bibinfo{booktitle}{\emph{Proceedings of the 18th ACM Conference on Computer Supported Cooperative Work \& Social Computing}} (Vancouver, BC, Canada) \emph{(\bibinfo{series}{CSCW '15})}. \bibinfo{publisher}{Association for Computing Machinery}, \bibinfo{address}{New York, NY, USA}, \bibinfo{pages}{1416–1426}.
\newblock
\showISBNx{9781450329224}
\urldef\tempurl%
\url{https://doi.org/10.1145/2675133.2675225}
\showDOI{\tempurl}


\bibitem[Daza and Ilozumba(2022)]%
        {Daza2022survey}
\bibfield{author}{\bibinfo{person}{Marco~Tulio Daza} {and} \bibinfo{person}{Usochi~Joanann Ilozumba}.} \bibinfo{year}{2022}\natexlab{}.
\newblock \showarticletitle{A survey of AI ethics in business literature: Maps and trends between 2000 and 2021}.
\newblock \bibinfo{journal}{\emph{Frontiers in Psychology}}  \bibinfo{volume}{13} (\bibinfo{year}{2022}), \bibinfo{pages}{1042661}.
\newblock
\urldef\tempurl%
\url{https://doi.org/10.3389/fpsyg.2022.1042661}
\showDOI{\tempurl}


\bibitem[Derner and Batisti{\v{c}}(2023)]%
        {derner2023beyond}
\bibfield{author}{\bibinfo{person}{Erik Derner} {and} \bibinfo{person}{Kristina Batisti{\v{c}}}.} \bibinfo{year}{2023}\natexlab{}.
\newblock \showarticletitle{Beyond the Safeguards: Exploring the Security Risks of ChatGPT}.
\newblock \bibinfo{journal}{\emph{arXiv preprint arXiv:2305.08005}} (\bibinfo{year}{2023}).
\newblock
\urldef\tempurl%
\url{https://doi.org/10.48550/arXiv.2305.08005}
\showDOI{\tempurl}


\bibitem[Devos-Comby and Devos(2001)]%
        {devos2001social}
\bibfield{author}{\bibinfo{person}{Loraine Devos-Comby} {and} \bibinfo{person}{Thierry Devos}.} \bibinfo{year}{2001}\natexlab{}.
\newblock \showarticletitle{Social norms, social value, and judgments of responsibility}.
\newblock \bibinfo{journal}{\emph{Swiss Journal of Psychology/Schweizerische Zeitschrift f{\"u}r Psychologie/Revue Suisse de Psychologie}} \bibinfo{volume}{60}, \bibinfo{number}{1} (\bibinfo{year}{2001}), \bibinfo{pages}{35}.
\newblock
\urldef\tempurl%
\url{https://doi.org/10.1024/1421-0185.60.1.35}
\showDOI{\tempurl}


\bibitem[Dignum(2017)]%
        {Dignum2017Responsible}
\bibfield{author}{\bibinfo{person}{Virginia Dignum}.} \bibinfo{year}{2017}\natexlab{}.
\newblock \showarticletitle{Responsible artificial intelligence: designing AI for human values}.
\newblock  (\bibinfo{year}{2017}).
\newblock
\urldef\tempurl%
\url{https://api.semanticscholar.org/CorpusID:158378636}
\showURL{%
\tempurl}


\bibitem[Dignum et~al\mbox{.}(2018)]%
        {Dignum2018Ethics}
\bibfield{author}{\bibinfo{person}{Virginia Dignum}, \bibinfo{person}{Matteo Baldoni}, \bibinfo{person}{Cristina Baroglio}, \bibinfo{person}{Maurizio Caon}, \bibinfo{person}{Raja Chatila}, \bibinfo{person}{Louise Dennis}, \bibinfo{person}{Gonzalo G{\'e}nova}, \bibinfo{person}{Galit Haim}, \bibinfo{person}{Malte~S Klie{\ss}}, \bibinfo{person}{Maite Lopez-Sanchez}, {et~al\mbox{.}}} \bibinfo{year}{2018}\natexlab{}.
\newblock \showarticletitle{Ethics by design: Necessity or curse?}. In \bibinfo{booktitle}{\emph{Proceedings of the 2018 AAAI/ACM Conference on AI, Ethics, and Society}}. \bibinfo{pages}{60--66}.
\newblock
\urldef\tempurl%
\url{https://doi.org/10.1145/3278721.3278745}
\showDOI{\tempurl}


\bibitem[Dwivedi et~al\mbox{.}(2023)]%
        {Dwivedi2023So}
\bibfield{author}{\bibinfo{person}{Yogesh~K Dwivedi}, \bibinfo{person}{Nir Kshetri}, \bibinfo{person}{Laurie Hughes}, \bibinfo{person}{Emma~Louise Slade}, \bibinfo{person}{Anand Jeyaraj}, \bibinfo{person}{Arpan~Kumar Kar}, \bibinfo{person}{Abdullah~M Baabdullah}, \bibinfo{person}{Alex Koohang}, \bibinfo{person}{Vishnupriya Raghavan}, \bibinfo{person}{Manju Ahuja}, {et~al\mbox{.}}} \bibinfo{year}{2023}\natexlab{}.
\newblock \showarticletitle{“So what if ChatGPT wrote it?” Multidisciplinary perspectives on opportunities, challenges and implications of generative conversational AI for research, practice and policy}.
\newblock \bibinfo{journal}{\emph{International Journal of Information Management}}  \bibinfo{volume}{71} (\bibinfo{year}{2023}), \bibinfo{pages}{102642}.
\newblock
\urldef\tempurl%
\url{https://doi.org/10.1016/j.ijinfomgt.2023.102642}
\showDOI{\tempurl}


\bibitem[Evans et~al\mbox{.}(2023)]%
        {evans2023chatgpt}
\bibfield{author}{\bibinfo{person}{Olaniyi Evans}, \bibinfo{person}{Olawale Wale-Awe}, \bibinfo{person}{Emeka Osuji}, \bibinfo{person}{Olawale Ayoola}, \bibinfo{person}{Raymond Alenoghena}, {and} \bibinfo{person}{Sesan Adeniji}.} \bibinfo{year}{2023}\natexlab{}.
\newblock \showarticletitle{ChatGPT impacts on access-efficiency, employment, education and ethics: The socio-economics of an AI language model}.
\newblock \bibinfo{journal}{\emph{BizEcons Quarterly}} \bibinfo{volume}{16}, \bibinfo{number}{1} (\bibinfo{year}{2023}), \bibinfo{pages}{1--17}.
\newblock


\bibitem[Farina et~al\mbox{.}(2024)]%
        {farina2024ai}
\bibfield{author}{\bibinfo{person}{Mirko Farina}, \bibinfo{person}{Petr Zhdanov}, \bibinfo{person}{Artur Karimov}, {and} \bibinfo{person}{Andrea Lavazza}.} \bibinfo{year}{2024}\natexlab{}.
\newblock \showarticletitle{AI and society: a virtue ethics approach}.
\newblock \bibinfo{journal}{\emph{AI \& SOCIETY}} \bibinfo{volume}{39}, \bibinfo{number}{3} (\bibinfo{year}{2024}), \bibinfo{pages}{1127--1140}.
\newblock


\bibitem[Fast and Horvitz(2017)]%
        {fast2017long}
\bibfield{author}{\bibinfo{person}{Ethan Fast} {and} \bibinfo{person}{Eric Horvitz}.} \bibinfo{year}{2017}\natexlab{}.
\newblock \showarticletitle{Long-term trends in the public perception of artificial intelligence}.
\newblock  \bibinfo{volume}{31}, \bibinfo{number}{1} (\bibinfo{year}{2017}).
\newblock
\urldef\tempurl%
\url{https://doi.org/10.1609/aaai.v31i1.10635}
\showDOI{\tempurl}


\bibitem[Fiesler et~al\mbox{.}(2018)]%
        {fiesler2018research}
\bibfield{author}{\bibinfo{person}{Casey Fiesler}, \bibinfo{person}{Amy Bruckman}, \bibinfo{person}{Robert~E Kraut}, \bibinfo{person}{Michael Muller}, \bibinfo{person}{Cosmin Munteanu}, {and} \bibinfo{person}{Katie Shilton}.} \bibinfo{year}{2018}\natexlab{}.
\newblock \showarticletitle{Research ethics and regulation: An open forum}. In \bibinfo{booktitle}{\emph{Companion of the 2018 ACM Conference on Computer Supported Cooperative Work and Social Computing}}. \bibinfo{pages}{125--128}.
\newblock


\bibitem[Fleischmann et~al\mbox{.}(2019)]%
        {10.1145/3311957.3359437}
\bibfield{author}{\bibinfo{person}{Kenneth~R. Fleischmann}, \bibinfo{person}{Sherri~R. Greenberg}, \bibinfo{person}{Danna Gurari}, \bibinfo{person}{Abigale Stangl}, \bibinfo{person}{Nitin Verma}, \bibinfo{person}{Jaxsen~R. Day}, \bibinfo{person}{Rachel~N. Simons}, {and} \bibinfo{person}{Tom Yeh}.} \bibinfo{year}{2019}\natexlab{}.
\newblock \showarticletitle{Good Systems: Ethical AI for CSCW}. In \bibinfo{booktitle}{\emph{Companion Publication of the 2019 Conference on Computer Supported Cooperative Work and Social Computing}} (Austin, TX, USA) \emph{(\bibinfo{series}{CSCW '19 Companion})}. \bibinfo{publisher}{Association for Computing Machinery}, \bibinfo{address}{New York, NY, USA}, \bibinfo{pages}{461–467}.
\newblock
\showISBNx{9781450366922}
\urldef\tempurl%
\url{https://doi.org/10.1145/3311957.3359437}
\showDOI{\tempurl}


\bibitem[Floridi(2019)]%
        {Floridi2019Translating}
\bibfield{author}{\bibinfo{person}{Luciano Floridi}.} \bibinfo{year}{2019}\natexlab{}.
\newblock \showarticletitle{Translating principles into practices of digital ethics: Five risks of being unethical}.
\newblock \bibinfo{journal}{\emph{Philosophy \& Technology}} \bibinfo{volume}{32}, \bibinfo{number}{2} (\bibinfo{year}{2019}), \bibinfo{pages}{185--193}.
\newblock
\urldef\tempurl%
\url{https://doi.org/10.1007/s13347-019-00354-x}
\showDOI{\tempurl}


\bibitem[Goertzel(2014)]%
        {goertzel2014artificial}
\bibfield{author}{\bibinfo{person}{Ben Goertzel}.} \bibinfo{year}{2014}\natexlab{}.
\newblock \showarticletitle{Artificial general intelligence: concept, state of the art, and future prospects}.
\newblock \bibinfo{journal}{\emph{Journal of Artificial General Intelligence}} \bibinfo{volume}{5}, \bibinfo{number}{1} (\bibinfo{year}{2014}), \bibinfo{pages}{1}.
\newblock
\urldef\tempurl%
\url{https://doi.org/10.2478/jagi-2014-0001}
\showDOI{\tempurl}


\bibitem[Gohel et~al\mbox{.}(2021)]%
        {Gohel2021Explainable}
\bibfield{author}{\bibinfo{person}{Prashant Gohel}, \bibinfo{person}{Priyanka Singh}, {and} \bibinfo{person}{Manoranjan Mohanty}.} \bibinfo{year}{2021}\natexlab{}.
\newblock \showarticletitle{Explainable AI: current status and future directions}.
\newblock \bibinfo{journal}{\emph{arXiv preprint arXiv:2107.07045}} (\bibinfo{year}{2021}).
\newblock
\urldef\tempurl%
\url{https://doi.org/10.48550/arXiv.2107.07045}
\showDOI{\tempurl}


\bibitem[Goldsmith and Burton(2017)]%
        {goldsmith2017teaching}
\bibfield{author}{\bibinfo{person}{Judy Goldsmith} {and} \bibinfo{person}{Emanuelle Burton}.} \bibinfo{year}{2017}\natexlab{}.
\newblock \showarticletitle{Why teaching ethics to AI practitioners is important}.
\newblock  \bibinfo{volume}{31}, \bibinfo{number}{1} (\bibinfo{year}{2017}).
\newblock
\urldef\tempurl%
\url{https://doi.org/10.1609/aaai.v31i1.11139}
\showDOI{\tempurl}


\bibitem[Govindarajulu et~al\mbox{.}(2019)]%
        {govindarajulu2019toward}
\bibfield{author}{\bibinfo{person}{Naveen~Sundar Govindarajulu}, \bibinfo{person}{Selmer Bringsjord}, \bibinfo{person}{Rikhiya Ghosh}, {and} \bibinfo{person}{Vasanth Sarathy}.} \bibinfo{year}{2019}\natexlab{}.
\newblock \showarticletitle{Toward the engineering of virtuous machines}. In \bibinfo{booktitle}{\emph{Proceedings of the 2019 AAAI/ACM Conference on AI, Ethics, and Society}}. \bibinfo{pages}{29--35}.
\newblock


\bibitem[Guarini(2012)]%
        {guarini2012conative}
\bibfield{author}{\bibinfo{person}{Marcello Guarini}.} \bibinfo{year}{2012}\natexlab{}.
\newblock \showarticletitle{Conative dimensions of machine ethics: A defense of duty}.
\newblock \bibinfo{journal}{\emph{IEEE Transactions on Affective Computing}} \bibinfo{volume}{3}, \bibinfo{number}{4} (\bibinfo{year}{2012}), \bibinfo{pages}{434--442}.
\newblock


\bibitem[Hagendorff(2020)]%
        {hagendorff2020ethics}
\bibfield{author}{\bibinfo{person}{Thilo Hagendorff}.} \bibinfo{year}{2020}\natexlab{}.
\newblock \showarticletitle{The ethics of AI ethics: An evaluation of guidelines}.
\newblock \bibinfo{journal}{\emph{Minds and machines}} \bibinfo{volume}{30}, \bibinfo{number}{1} (\bibinfo{year}{2020}), \bibinfo{pages}{99--120}.
\newblock
\urldef\tempurl%
\url{https://doi.org/10.1007/s11023-020-09517-8}
\showDOI{\tempurl}


\bibitem[Hagendorff(2022)]%
        {Hagendorff2022virtue}
\bibfield{author}{\bibinfo{person}{Thilo Hagendorff}.} \bibinfo{year}{2022}\natexlab{}.
\newblock \showarticletitle{A virtue-based framework to support putting AI ethics into practice}.
\newblock \bibinfo{journal}{\emph{Philosophy \& Technology}} \bibinfo{volume}{35}, \bibinfo{number}{3} (\bibinfo{year}{2022}), \bibinfo{pages}{55}.
\newblock
\urldef\tempurl%
\url{https://doi.org/10.1007/s13347-022-00553-z}
\showDOI{\tempurl}


\bibitem[Hallamaa and Kalliokoski(2022)]%
        {Hallamaa2022AI}
\bibfield{author}{\bibinfo{person}{Jaana Hallamaa} {and} \bibinfo{person}{Taina Kalliokoski}.} \bibinfo{year}{2022}\natexlab{}.
\newblock \showarticletitle{AI ethics as applied ethics}.
\newblock \bibinfo{journal}{\emph{Frontiers in computer science}}  \bibinfo{volume}{4} (\bibinfo{year}{2022}), \bibinfo{pages}{776837}.
\newblock
\urldef\tempurl%
\url{https://doi.org/10.3389/fcomp.2022.776837}
\showDOI{\tempurl}


\bibitem[Hasal et~al\mbox{.}(2021)]%
        {Hasal2021Chatbots}
\bibfield{author}{\bibinfo{person}{Martin Hasal}, \bibinfo{person}{Jana Nowakov{\'a}}, \bibinfo{person}{Khalifa Ahmed~Saghair}, \bibinfo{person}{Hussam Abdulla}, \bibinfo{person}{V{\'a}clav Sn{\'a}{\v{s}}el}, {and} \bibinfo{person}{Lidia Ogiela}.} \bibinfo{year}{2021}\natexlab{}.
\newblock \showarticletitle{Chatbots: Security, privacy, data protection, and social aspects}.
\newblock \bibinfo{journal}{\emph{Concurrency and Computation: Practice and Experience}} \bibinfo{volume}{33}, \bibinfo{number}{19} (\bibinfo{year}{2021}), \bibinfo{pages}{e6426}.
\newblock
\urldef\tempurl%
\url{https://doi.org/10.1002/cpe.6426}
\showDOI{\tempurl}


\bibitem[Hauptman et~al\mbox{.}(2022)]%
        {10.1145/3500868.3559450}
\bibfield{author}{\bibinfo{person}{Allyson~I. Hauptman}, \bibinfo{person}{Wen Duan}, {and} \bibinfo{person}{Nathan~J. Mcneese}.} \bibinfo{year}{2022}\natexlab{}.
\newblock \showarticletitle{The Components of Trust for Collaborating With AI Colleagues}. In \bibinfo{booktitle}{\emph{Companion Publication of the 2022 Conference on Computer Supported Cooperative Work and Social Computing}} (Virtual Event, Taiwan) \emph{(\bibinfo{series}{CSCW'22 Companion})}. \bibinfo{publisher}{Association for Computing Machinery}, \bibinfo{address}{New York, NY, USA}, \bibinfo{pages}{72–75}.
\newblock
\showISBNx{9781450391900}
\urldef\tempurl%
\url{https://doi.org/10.1145/3500868.3559450}
\showDOI{\tempurl}


\bibitem[Hausman(2008)]%
        {hausman2008fairness}
\bibfield{author}{\bibinfo{person}{Daniel~M Hausman}.} \bibinfo{year}{2008}\natexlab{}.
\newblock \showarticletitle{Fairness and social norms}.
\newblock \bibinfo{journal}{\emph{Philosophy of science}} \bibinfo{volume}{75}, \bibinfo{number}{5} (\bibinfo{year}{2008}), \bibinfo{pages}{850--860}.
\newblock
\urldef\tempurl%
\url{https://doi.org/10.1086/594529}
\showDOI{\tempurl}


\bibitem[Hermann(2022)]%
        {hermann2022leveraging}
\bibfield{author}{\bibinfo{person}{Erik Hermann}.} \bibinfo{year}{2022}\natexlab{}.
\newblock \showarticletitle{Leveraging artificial intelligence in marketing for social good—An ethical perspective}.
\newblock \bibinfo{journal}{\emph{Journal of Business Ethics}} \bibinfo{volume}{179}, \bibinfo{number}{1} (\bibinfo{year}{2022}), \bibinfo{pages}{43--61}.
\newblock
\urldef\tempurl%
\url{https://doi.org/10.1007/s10551-021-04843-y}
\showDOI{\tempurl}


\bibitem[Hettiachchi et~al\mbox{.}(2021)]%
        {10.1145/3462204.3481729}
\bibfield{author}{\bibinfo{person}{Danula Hettiachchi}, \bibinfo{person}{Mark Sanderson}, \bibinfo{person}{Jorge Goncalves}, \bibinfo{person}{Simo Hosio}, \bibinfo{person}{Gabriella Kazai}, \bibinfo{person}{Matthew Lease}, \bibinfo{person}{Mike Schaekermann}, {and} \bibinfo{person}{Emine Yilmaz}.} \bibinfo{year}{2021}\natexlab{}.
\newblock \showarticletitle{Investigating and Mitigating Biases in Crowdsourced Data}. In \bibinfo{booktitle}{\emph{Companion Publication of the 2021 Conference on Computer Supported Cooperative Work and Social Computing}} (Virtual Event, USA) \emph{(\bibinfo{series}{CSCW '21 Companion})}. \bibinfo{publisher}{Association for Computing Machinery}, \bibinfo{address}{New York, NY, USA}, \bibinfo{pages}{331–334}.
\newblock
\showISBNx{9781450384797}
\urldef\tempurl%
\url{https://doi.org/10.1145/3462204.3481729}
\showDOI{\tempurl}


\bibitem[Hu and Leung(2018)]%
        {hu2018social}
\bibfield{author}{\bibinfo{person}{Shuyue Hu} {and} \bibinfo{person}{Ho-fung Leung}.} \bibinfo{year}{2018}\natexlab{}.
\newblock \showarticletitle{Do Social Norms Emerge? The Evolution of Agents' Decisions with the Awareness of Social Values under Iterated Prisoner's Dilemma}. In \bibinfo{booktitle}{\emph{2018 IEEE 12th International Conference on Self-Adaptive and Self-Organizing Systems (SASO)}}. IEEE, \bibinfo{pages}{11--19}.
\newblock
\urldef\tempurl%
\url{https://doi.org/10.1109/SASO.2018.00012}
\showDOI{\tempurl}


\bibitem[Huang et~al\mbox{.}(2023)]%
        {Huang2023IS}
\bibfield{author}{\bibinfo{person}{Fan Huang}, \bibinfo{person}{Haewoon Kwak}, {and} \bibinfo{person}{Jisun An}.} \bibinfo{year}{2023}\natexlab{}.
\newblock \showarticletitle{Is chatgpt better than human annotators? potential and limitations of chatgpt in explaining implicit hate speech}.
\newblock \bibinfo{journal}{\emph{arXiv preprint arXiv:2302.07736}} (\bibinfo{year}{2023}).
\newblock
\urldef\tempurl%
\url{https://doi.org/10.48550/arXiv.2302.07736}
\showDOI{\tempurl}


\bibitem[Huang et~al\mbox{.}(2024)]%
        {huang2024survey}
\bibfield{author}{\bibinfo{person}{Xiaowei Huang}, \bibinfo{person}{Wenjie Ruan}, \bibinfo{person}{Wei Huang}, \bibinfo{person}{Gaojie Jin}, \bibinfo{person}{Yi Dong}, \bibinfo{person}{Changshun Wu}, \bibinfo{person}{Saddek Bensalem}, \bibinfo{person}{Ronghui Mu}, \bibinfo{person}{Yi Qi}, \bibinfo{person}{Xingyu Zhao}, {et~al\mbox{.}}} \bibinfo{year}{2024}\natexlab{}.
\newblock \showarticletitle{A survey of safety and trustworthiness of large language models through the lens of verification and validation}.
\newblock \bibinfo{journal}{\emph{Artificial Intelligence Review}} \bibinfo{volume}{57}, \bibinfo{number}{7} (\bibinfo{year}{2024}), \bibinfo{pages}{175}.
\newblock


\bibitem[Jobin et~al\mbox{.}(2019)]%
        {jobin2019global}
\bibfield{author}{\bibinfo{person}{Anna Jobin}, \bibinfo{person}{Marcello Ienca}, {and} \bibinfo{person}{Effy Vayena}.} \bibinfo{year}{2019}\natexlab{}.
\newblock \showarticletitle{The global landscape of AI ethics guidelines}.
\newblock \bibinfo{journal}{\emph{Nature machine intelligence}} \bibinfo{volume}{1}, \bibinfo{number}{9} (\bibinfo{year}{2019}), \bibinfo{pages}{389--399}.
\newblock
\urldef\tempurl%
\url{https://doi.org/10.1038/s42256-019-0088-2}
\showDOI{\tempurl}


\bibitem[Kulkarni et~al\mbox{.}(2023)]%
        {10.1145/3584931.3608438}
\bibfield{author}{\bibinfo{person}{Chinmay Kulkarni}, \bibinfo{person}{Tongshuang Wu}, \bibinfo{person}{Kenneth Holstein}, \bibinfo{person}{Q.~Vera Liao}, \bibinfo{person}{Min~Kyung Lee}, \bibinfo{person}{Mina Lee}, {and} \bibinfo{person}{Hariharan Subramonyam}.} \bibinfo{year}{2023}\natexlab{}.
\newblock \showarticletitle{LLMs and the Infrastructure of CSCW}. In \bibinfo{booktitle}{\emph{Companion Publication of the 2023 Conference on Computer Supported Cooperative Work and Social Computing}} (Minneapolis, MN, USA) \emph{(\bibinfo{series}{CSCW '23 Companion})}. \bibinfo{publisher}{Association for Computing Machinery}, \bibinfo{address}{New York, NY, USA}, \bibinfo{pages}{408–410}.
\newblock
\showISBNx{9798400701290}
\urldef\tempurl%
\url{https://doi.org/10.1145/3584931.3608438}
\showDOI{\tempurl}


\bibitem[LaCroix and Bengio(2019)]%
        {LaCroix2019Learning}
\bibfield{author}{\bibinfo{person}{Travis LaCroix} {and} \bibinfo{person}{Yoshua Bengio}.} \bibinfo{year}{2019}\natexlab{}.
\newblock \showarticletitle{Learning from learning machines: optimisation, rules, and social norms}.
\newblock \bibinfo{journal}{\emph{arXiv preprint arXiv:2001.00006}} (\bibinfo{year}{2019}).
\newblock
\urldef\tempurl%
\url{https://doi.org/10.48550/arXiv.2001.00006}
\showDOI{\tempurl}


\bibitem[Lagerkvist et~al\mbox{.}(2024)]%
        {lagerkvist2024body}
\bibfield{author}{\bibinfo{person}{Amanda Lagerkvist}, \bibinfo{person}{Matilda Tudor}, \bibinfo{person}{Jacek Smolicki}, \bibinfo{person}{Charles~M Ess}, \bibinfo{person}{Jenny Eriksson~Lundstr{\"o}m}, {and} \bibinfo{person}{Maria Rogg}.} \bibinfo{year}{2024}\natexlab{}.
\newblock \showarticletitle{Body stakes: an existential ethics of care in living with biometrics and AI}.
\newblock \bibinfo{journal}{\emph{AI \& SOCIETY}} \bibinfo{volume}{39}, \bibinfo{number}{1} (\bibinfo{year}{2024}), \bibinfo{pages}{169--181}.
\newblock


\bibitem[Lauer(2021)]%
        {Lauer2020cannot}
\bibfield{author}{\bibinfo{person}{Dave Lauer}.} \bibinfo{year}{2021}\natexlab{}.
\newblock \showarticletitle{You cannot have AI ethics without ethics}.
\newblock \bibinfo{journal}{\emph{AI and Ethics}} \bibinfo{volume}{1}, \bibinfo{number}{1} (\bibinfo{year}{2021}), \bibinfo{pages}{21--25}.
\newblock
\urldef\tempurl%
\url{https://doi.org/10.1007/s43681-020-00013-4}
\showDOI{\tempurl}


\bibitem[Lee et~al\mbox{.}(2024)]%
        {lee2024one}
\bibfield{author}{\bibinfo{person}{Yoonjoo Lee}, \bibinfo{person}{Kihoon Son}, \bibinfo{person}{Tae~Soo Kim}, \bibinfo{person}{Jisu Kim}, \bibinfo{person}{John Joon~Young Chung}, \bibinfo{person}{Eytan Adar}, {and} \bibinfo{person}{Juho Kim}.} \bibinfo{year}{2024}\natexlab{}.
\newblock \showarticletitle{One vs. Many: Comprehending Accurate Information from Multiple Erroneous and Inconsistent AI Generations}. In \bibinfo{booktitle}{\emph{The 2024 ACM Conference on Fairness, Accountability, and Transparency}}. \bibinfo{pages}{2518--2531}.
\newblock


\bibitem[Leslie(2019)]%
        {Leslie2019Understanding}
\bibfield{author}{\bibinfo{person}{David Leslie}.} \bibinfo{year}{2019}\natexlab{}.
\newblock \showarticletitle{Understanding artificial intelligence ethics and safety}.
\newblock \bibinfo{journal}{\emph{arXiv preprint arXiv:1906.05684}} (\bibinfo{year}{2019}).
\newblock
\urldef\tempurl%
\url{https://www.turing.ac.uk/news/publications/understanding-artificial-intelligence-ethics-and-safety}
\showURL{%
\tempurl}


\bibitem[Li et~al\mbox{.}(2022)]%
        {li2022ethical}
\bibfield{author}{\bibinfo{person}{Hanlin Li}, \bibinfo{person}{Leah Ajmani}, \bibinfo{person}{Moyan Zhou}, \bibinfo{person}{Nicholas Vincent}, \bibinfo{person}{Sohyeon Hwang}, \bibinfo{person}{Tiziano Piccardi}, \bibinfo{person}{Sneha Narayan}, \bibinfo{person}{Sherae Daniel}, {and} \bibinfo{person}{Veniamin Veselovsky}.} \bibinfo{year}{2022}\natexlab{}.
\newblock \showarticletitle{Ethical tensions, norms, and directions in the extraction of online volunteer work}. In \bibinfo{booktitle}{\emph{Companion Publication of the 2022 Conference on Computer Supported Cooperative Work and Social Computing}}. \bibinfo{pages}{273--277}.
\newblock


\bibitem[Liang et~al\mbox{.}(2021)]%
        {liang2021towards}
\bibfield{author}{\bibinfo{person}{Paul~Pu Liang}, \bibinfo{person}{Chiyu Wu}, \bibinfo{person}{Louis-Philippe Morency}, {and} \bibinfo{person}{Ruslan Salakhutdinov}.} \bibinfo{year}{2021}\natexlab{}.
\newblock \showarticletitle{Towards understanding and mitigating social biases in language models}. In \bibinfo{booktitle}{\emph{International Conference on Machine Learning}}. PMLR, \bibinfo{pages}{6565--6576}.
\newblock


\bibitem[Lima et~al\mbox{.}(2021)]%
        {lima2021human}
\bibfield{author}{\bibinfo{person}{Gabriel Lima}, \bibinfo{person}{Nina Grgi{\'c}-Hla{\v{c}}a}, {and} \bibinfo{person}{Meeyoung Cha}.} \bibinfo{year}{2021}\natexlab{}.
\newblock \showarticletitle{Human perceptions on moral responsibility of AI: A case study in AI-assisted bail decision-making}. In \bibinfo{booktitle}{\emph{Proceedings of the 2021 CHI Conference on Human Factors in Computing Systems}}. \bibinfo{pages}{1--17}.
\newblock
\urldef\tempurl%
\url{https://doi.org/10.1145/3411764.3445260}
\showDOI{\tempurl}


\bibitem[Loi et~al\mbox{.}(2019)]%
        {10.1145/3301019.3320000}
\bibfield{author}{\bibinfo{person}{Daria Loi}, \bibinfo{person}{Christine~T. Wolf}, \bibinfo{person}{Jeanette~L. Blomberg}, \bibinfo{person}{Raphael Arar}, {and} \bibinfo{person}{Margot Brereton}.} \bibinfo{year}{2019}\natexlab{}.
\newblock \showarticletitle{Co-Designing AI Futures: Integrating AI Ethics, Social Computing, and Design}. In \bibinfo{booktitle}{\emph{Companion Publication of the 2019 on Designing Interactive Systems Conference 2019 Companion}} \emph{(\bibinfo{series}{DIS '19 Companion})}. \bibinfo{publisher}{Association for Computing Machinery}, \bibinfo{address}{New York, NY, USA}, \bibinfo{pages}{381–384}.
\newblock
\showISBNx{9781450362702}
\urldef\tempurl%
\url{https://doi.org/10.1145/3301019.3320000}
\showDOI{\tempurl}


\bibitem[Madaio et~al\mbox{.}(2020)]%
        {madaio2020co}
\bibfield{author}{\bibinfo{person}{Michael~A. Madaio}, \bibinfo{person}{Luke Stark}, \bibinfo{person}{Jennifer Wortman~Vaughan}, {and} \bibinfo{person}{Hanna Wallach}.} \bibinfo{year}{2020}\natexlab{}.
\newblock \showarticletitle{Co-Designing Checklists to Understand Organizational Challenges and Opportunities around Fairness in AI}. In \bibinfo{booktitle}{\emph{Proceedings of the 2020 CHI Conference on Human Factors in Computing Systems}} (Honolulu, HI, USA) \emph{(\bibinfo{series}{CHI '20})}. \bibinfo{publisher}{Association for Computing Machinery}, \bibinfo{address}{New York, NY, USA}, \bibinfo{pages}{1–14}.
\newblock
\showISBNx{9781450367080}
\urldef\tempurl%
\url{https://doi.org/10.1145/3313831.3376445}
\showDOI{\tempurl}


\bibitem[Maitra(2020)]%
        {maitra2020artificial}
\bibfield{author}{\bibinfo{person}{Suvradip Maitra}.} \bibinfo{year}{2020}\natexlab{}.
\newblock \showarticletitle{Artificial intelligence and indigenous perspectives: Protecting and empowering intelligent human beings}. In \bibinfo{booktitle}{\emph{Proceedings of the AAAI/ACM Conference on AI, Ethics, and Society}}. \bibinfo{pages}{320--326}.
\newblock


\bibitem[McGee(2023)]%
        {McGee2023IS}
\bibfield{author}{\bibinfo{person}{Robert~W McGee}.} \bibinfo{year}{2023}\natexlab{}.
\newblock \showarticletitle{Is chat gpt biased against conservatives? an empirical study}.
\newblock \bibinfo{journal}{\emph{An Empirical Study (February 15, 2023)}} (\bibinfo{year}{2023}).
\newblock
\urldef\tempurl%
\url{https://doi.org/10.2139/ssrn.4359405}
\showDOI{\tempurl}


\bibitem[McKight and Najab(2010)]%
        {mckight2010kruskal}
\bibfield{author}{\bibinfo{person}{Patrick~E McKight} {and} \bibinfo{person}{Julius Najab}.} \bibinfo{year}{2010}\natexlab{}.
\newblock \showarticletitle{Kruskal-wallis test}.
\newblock \bibinfo{journal}{\emph{The corsini encyclopedia of psychology}} (\bibinfo{year}{2010}), \bibinfo{pages}{1--1}.
\newblock


\bibitem[Mhlanga(2023)]%
        {mhlanga2023open}
\bibfield{author}{\bibinfo{person}{David Mhlanga}.} \bibinfo{year}{2023}\natexlab{}.
\newblock \showarticletitle{Open AI in education, the responsible and ethical use of ChatGPT towards lifelong learning}.
\newblock \bibinfo{journal}{\emph{Education, the Responsible and Ethical Use of ChatGPT Towards Lifelong Learning (February 11, 2023)}} (\bibinfo{year}{2023}).
\newblock
\urldef\tempurl%
\url{https://doi.org/10.2139/ssrn.4354422}
\showDOI{\tempurl}


\bibitem[Mijwil et~al\mbox{.}(2023)]%
        {Mijwil2023ChatGPT}
\bibfield{author}{\bibinfo{person}{Maad~M Mijwil}, \bibinfo{person}{Kamal~Kant Hiran}, \bibinfo{person}{Ruchi Doshi}, \bibinfo{person}{Manish Dadhich}, \bibinfo{person}{Abdel-Hameed Al-Mistarehi}, {and} \bibinfo{person}{Indu Bala}.} \bibinfo{year}{2023}\natexlab{}.
\newblock \showarticletitle{ChatGPT and the future of academic integrity in the artificial intelligence era: a new frontier}.
\newblock \bibinfo{journal}{\emph{Al-Salam Journal for Engineering and Technology}} \bibinfo{volume}{2}, \bibinfo{number}{2} (\bibinfo{year}{2023}), \bibinfo{pages}{116--127}.
\newblock
\urldef\tempurl%
\url{https://doi.org/10.55145/ajest.2023.02.02.015}
\showDOI{\tempurl}


\bibitem[Moor(1995)]%
        {Moor1995Is}
\bibfield{author}{\bibinfo{person}{James~H Moor}.} \bibinfo{year}{1995}\natexlab{}.
\newblock \showarticletitle{Is ethics computable?}
\newblock \bibinfo{journal}{\emph{Metaphilosophy}} \bibinfo{volume}{26}, \bibinfo{number}{1/2} (\bibinfo{year}{1995}), \bibinfo{pages}{1--21}.
\newblock
\urldef\tempurl%
\url{https://www.jstor.org/stable/24439044}
\showURL{%
\tempurl}


\bibitem[Morley et~al\mbox{.}(2020)]%
        {Morley2020The}
\bibfield{author}{\bibinfo{person}{Jessica Morley}, \bibinfo{person}{Caio~CV Machado}, \bibinfo{person}{Christopher Burr}, \bibinfo{person}{Josh Cowls}, \bibinfo{person}{Indra Joshi}, \bibinfo{person}{Mariarosaria Taddeo}, {and} \bibinfo{person}{Luciano Floridi}.} \bibinfo{year}{2020}\natexlab{}.
\newblock \showarticletitle{The ethics of AI in health care: a mapping review}.
\newblock \bibinfo{journal}{\emph{Social Science \& Medicine}}  \bibinfo{volume}{260} (\bibinfo{year}{2020}), \bibinfo{pages}{113172}.
\newblock
\urldef\tempurl%
\url{https://doi.org/10.1016/j.socscimed.2020.113172}
\showDOI{\tempurl}


\bibitem[Muyskens et~al\mbox{.}(2024)]%
        {muyskens2024can}
\bibfield{author}{\bibinfo{person}{Kathryn Muyskens}, \bibinfo{person}{Yonghui Ma}, {and} \bibinfo{person}{Michael Dunn}.} \bibinfo{year}{2024}\natexlab{}.
\newblock \showarticletitle{Can an AI-carebot be filial? Reflections from Confucian ethics}.
\newblock \bibinfo{journal}{\emph{Nursing Ethics}} (\bibinfo{year}{2024}), \bibinfo{pages}{09697330241238332}.
\newblock


\bibitem[Neuman and Cohen(2023)]%
        {neuman2023ai}
\bibfield{author}{\bibinfo{person}{Yair Neuman} {and} \bibinfo{person}{Yochai Cohen}.} \bibinfo{year}{2023}\natexlab{}.
\newblock \showarticletitle{AI for identifying social norm violation}.
\newblock \bibinfo{journal}{\emph{Scientific Reports}} \bibinfo{volume}{13}, \bibinfo{number}{1} (\bibinfo{year}{2023}), \bibinfo{pages}{8103}.
\newblock
\urldef\tempurl%
\url{https://doi.org/s41598-023-35350-x}
\showDOI{\tempurl}


\bibitem[Oviedo-Trespalacios et~al\mbox{.}({[n.\,d.]})]%
        {Oviedo2023The}
\bibfield{author}{\bibinfo{person}{O Oviedo-Trespalacios}, \bibinfo{person}{AE Peden}, \bibinfo{person}{T Cole-Hunter}, \bibinfo{person}{A Costantini}, \bibinfo{person}{M Haghani}, \bibinfo{person}{S Kelly}, {and} \bibinfo{person}{G Reniers}.} \bibinfo{year}{[n.\,d.]}\natexlab{}.
\newblock \showarticletitle{The risks of using ChatGPT to obtain common safety-related information and advice, 2023}.
\newblock \bibinfo{journal}{\emph{Available At SSRN}}  \bibinfo{volume}{4346827} (\bibinfo{year}{[n.\,d.]}).
\newblock
\urldef\tempurl%
\url{https://doi.org/10.1016/j.ssci.2023.106244}
\showDOI{\tempurl}


\bibitem[Poller et~al\mbox{.}(2017)]%
        {10.1145/2998181.2998191}
\bibfield{author}{\bibinfo{person}{Andreas Poller}, \bibinfo{person}{Laura Kocksch}, \bibinfo{person}{Sven T\"{u}rpe}, \bibinfo{person}{Felix~Anand Epp}, {and} \bibinfo{person}{Katharina Kinder-Kurlanda}.} \bibinfo{year}{2017}\natexlab{}.
\newblock \showarticletitle{Can Security Become a Routine? A Study of Organizational Change in an Agile Software Development Group}. In \bibinfo{booktitle}{\emph{Proceedings of the 2017 ACM Conference on Computer Supported Cooperative Work and Social Computing}} (Portland, Oregon, USA) \emph{(\bibinfo{series}{CSCW '17})}. \bibinfo{publisher}{Association for Computing Machinery}, \bibinfo{address}{New York, NY, USA}, \bibinfo{pages}{2489–2503}.
\newblock
\showISBNx{9781450343350}
\urldef\tempurl%
\url{https://doi.org/10.1145/2998181.2998191}
\showDOI{\tempurl}


\bibitem[Poslon and {\v{C}}artolovni({[n.\,d.]})]%
        {poslonoutpacing}
\bibfield{author}{\bibinfo{person}{Luka Poslon} {and} \bibinfo{person}{Anto {\v{C}}artolovni}.} \bibinfo{year}{[n.\,d.]}\natexlab{}.
\newblock \showarticletitle{Outpacing the Trustworthiness in LLM Use in Medicine by Addressing Opacity and Enhancing Explainability}.
\newblock \bibinfo{journal}{\emph{Frontiers of Artificial Intelligence—Philosophical Explorations}} (\bibinfo{year}{[n.\,d.]}).
\newblock


\bibitem[Quttainah et~al\mbox{.}(2024)]%
        {quttainah2024cost}
\bibfield{author}{\bibinfo{person}{Majdi Quttainah}, \bibinfo{person}{Vinaytosh Mishra}, \bibinfo{person}{Somayya Madakam}, \bibinfo{person}{Yotam Lurie}, \bibinfo{person}{Shlomo Mark}, {et~al\mbox{.}}} \bibinfo{year}{2024}\natexlab{}.
\newblock \showarticletitle{Cost, Usability, Credibility, Fairness, Accountability, Transparency, and Explainability Framework for Safe and Effective Large Language Models in Medical Education: Narrative Review and Qualitative Study}.
\newblock \bibinfo{journal}{\emph{JMIR AI}} \bibinfo{volume}{3}, \bibinfo{number}{1} (\bibinfo{year}{2024}), \bibinfo{pages}{e51834}.
\newblock


\bibitem[Ray(2023)]%
        {Ray2023ChatGPT}
\bibfield{author}{\bibinfo{person}{Partha~Pratim Ray}.} \bibinfo{year}{2023}\natexlab{}.
\newblock \showarticletitle{ChatGPT: A comprehensive review on background, applications, key challenges, bias, ethics, limitations and future scope}.
\newblock \bibinfo{journal}{\emph{Internet of Things and Cyber-Physical Systems}} (\bibinfo{year}{2023}).
\newblock
\urldef\tempurl%
\url{https://doi.org/10.1016/j.iotcps.2023.04.003}
\showDOI{\tempurl}


\bibitem[Ress{\'e}guier and Rodrigues(2020)]%
        {Resseguier2020ethics}
\bibfield{author}{\bibinfo{person}{Ana{\"\i}s Ress{\'e}guier} {and} \bibinfo{person}{Rowena Rodrigues}.} \bibinfo{year}{2020}\natexlab{}.
\newblock \showarticletitle{AI ethics should not remain toothless! A call to bring back the teeth of ethics}.
\newblock \bibinfo{journal}{\emph{Big Data \& Society}} \bibinfo{volume}{7}, \bibinfo{number}{2} (\bibinfo{year}{2020}), \bibinfo{pages}{2053951720942541}.
\newblock
\urldef\tempurl%
\url{https://doi.org/10.1177/2053951720942541}
\showDOI{\tempurl}


\bibitem[Rivas and Zhao(2023)]%
        {rivas2023marketing}
\bibfield{author}{\bibinfo{person}{Pablo Rivas} {and} \bibinfo{person}{Liang Zhao}.} \bibinfo{year}{2023}\natexlab{}.
\newblock \showarticletitle{Marketing with chatgpt: Navigating the ethical terrain of gpt-based chatbot technology}.
\newblock \bibinfo{journal}{\emph{AI}} \bibinfo{volume}{4}, \bibinfo{number}{2} (\bibinfo{year}{2023}), \bibinfo{pages}{375--384}.
\newblock
\urldef\tempurl%
\url{https://doi.org/10.3390/ai4020019}
\showDOI{\tempurl}


\bibitem[Roberts and Lee(2014)]%
        {roberts2014autonomy}
\bibfield{author}{\bibinfo{person}{Lisa~R Roberts} {and} \bibinfo{person}{Jerry~W Lee}.} \bibinfo{year}{2014}\natexlab{}.
\newblock \showarticletitle{Autonomy and social norms in a three factor grief model predicting perinatal grief in India}.
\newblock \bibinfo{journal}{\emph{Health Care for Women International}} \bibinfo{volume}{35}, \bibinfo{number}{3} (\bibinfo{year}{2014}), \bibinfo{pages}{285--299}.
\newblock


\bibitem[Rossi and Mattei(2019)]%
        {rossi2019building}
\bibfield{author}{\bibinfo{person}{Francesca Rossi} {and} \bibinfo{person}{Nicholas Mattei}.} \bibinfo{year}{2019}\natexlab{}.
\newblock \showarticletitle{Building ethically bounded AI}. In \bibinfo{booktitle}{\emph{Proceedings of the AAAI Conference on Artificial Intelligence}}, Vol.~\bibinfo{volume}{33}. \bibinfo{pages}{9785--9789}.
\newblock
\urldef\tempurl%
\url{https://doi.org/10.1609/aaai.v33i01.33019785}
\showDOI{\tempurl}


\bibitem[Ryan and Stahl(2020)]%
        {Ryan2020Artificial}
\bibfield{author}{\bibinfo{person}{Mark Ryan} {and} \bibinfo{person}{Bernd~Carsten Stahl}.} \bibinfo{year}{2020}\natexlab{}.
\newblock \showarticletitle{Artificial intelligence ethics guidelines for developers and users: clarifying their content and normative implications}.
\newblock \bibinfo{journal}{\emph{Journal of Information, Communication and Ethics in Society}} \bibinfo{volume}{19}, \bibinfo{number}{1} (\bibinfo{year}{2020}), \bibinfo{pages}{61--86}.
\newblock
\urldef\tempurl%
\url{https://doi.org/10.1108/jices-12-2019-0138}
\showDOI{\tempurl}


\bibitem[Sag(2023)]%
        {sag2023copyright}
\bibfield{author}{\bibinfo{person}{Matthew Sag}.} \bibinfo{year}{2023}\natexlab{}.
\newblock \showarticletitle{Copyright safety for generative ai}.
\newblock \bibinfo{journal}{\emph{Hous. L. Rev.}}  \bibinfo{volume}{61} (\bibinfo{year}{2023}), \bibinfo{pages}{295}.
\newblock


\bibitem[Salo-P{\"o}ntinen(2021)]%
        {Salo2021AI}
\bibfield{author}{\bibinfo{person}{Henrikki Salo-P{\"o}ntinen}.} \bibinfo{year}{2021}\natexlab{}.
\newblock \showarticletitle{AI Ethics-Critical Reflections on Embedding Ethical Frameworks in AI Technology}. In \bibinfo{booktitle}{\emph{International Conference on Human-Computer Interaction}}. Springer, \bibinfo{pages}{311--329}.
\newblock
\urldef\tempurl%
\url{https://doi.org/10.1007/978-3-030-77431-8_20}
\showDOI{\tempurl}


\bibitem[Samarawickrama(2022)]%
        {Samarawickrama2022AI}
\bibfield{author}{\bibinfo{person}{Mahendra Samarawickrama}.} \bibinfo{year}{2022}\natexlab{}.
\newblock \showarticletitle{AI Governance and Ethics Framework for Sustainable AI and Sustainability}.
\newblock \bibinfo{journal}{\emph{arXiv preprint arXiv:2210.08984}} (\bibinfo{year}{2022}).
\newblock
\urldef\tempurl%
\url{https://doi.org/10.48550/arXiv.2210.08984}
\showDOI{\tempurl}


\bibitem[Samuelson(2023)]%
        {samuelson2023generative}
\bibfield{author}{\bibinfo{person}{Pamela Samuelson}.} \bibinfo{year}{2023}\natexlab{}.
\newblock \showarticletitle{Generative AI meets copyright}.
\newblock \bibinfo{journal}{\emph{Science}} \bibinfo{volume}{381}, \bibinfo{number}{6654} (\bibinfo{year}{2023}), \bibinfo{pages}{158--161}.
\newblock


\bibitem[Santos et~al\mbox{.}(2018)]%
        {santos2018social}
\bibfield{author}{\bibinfo{person}{Fernando Santos}, \bibinfo{person}{Jorge Pacheco}, {and} \bibinfo{person}{Francisco Santos}.} \bibinfo{year}{2018}\natexlab{}.
\newblock \showarticletitle{Social norms of cooperation with costly reputation building}. In \bibinfo{booktitle}{\emph{Proceedings of the AAAI Conference on Artificial Intelligence}}, Vol.~\bibinfo{volume}{32}.
\newblock
\urldef\tempurl%
\url{https://doi.org/10.1609/aaai.v32i1.11582}
\showDOI{\tempurl}


\bibitem[Schlesinger et~al\mbox{.}(2018)]%
        {schlesinger2018let}
\bibfield{author}{\bibinfo{person}{Ari Schlesinger}, \bibinfo{person}{Kenton~P O'Hara}, {and} \bibinfo{person}{Alex~S Taylor}.} \bibinfo{year}{2018}\natexlab{}.
\newblock \showarticletitle{Let's talk about race: Identity, chatbots, and AI}. In \bibinfo{booktitle}{\emph{Proceedings of the 2018 chi conference on human factors in computing systems}}. \bibinfo{pages}{1--14}.
\newblock
\urldef\tempurl%
\url{https://doi.org/10.1145/3173574.3173889}
\showDOI{\tempurl}


\bibitem[Shen et~al\mbox{.}(2023b)]%
        {10.1145/3584931.3607492}
\bibfield{author}{\bibinfo{person}{Hua Shen}, \bibinfo{person}{Chieh-Yang Huang}, \bibinfo{person}{Tongshuang Wu}, {and} \bibinfo{person}{Ting-Hao~Kenneth Huang}.} \bibinfo{year}{2023}\natexlab{b}.
\newblock \showarticletitle{ConvXAI: Delivering Heterogeneous AI Explanations via Conversations to Support Human-AI Scientific Writing}. In \bibinfo{booktitle}{\emph{Companion Publication of the 2023 Conference on Computer Supported Cooperative Work and Social Computing}} (Minneapolis, MN, USA) \emph{(\bibinfo{series}{CSCW '23 Companion})}. \bibinfo{publisher}{Association for Computing Machinery}, \bibinfo{address}{New York, NY, USA}, \bibinfo{pages}{384–387}.
\newblock
\showISBNx{9798400701290}
\urldef\tempurl%
\url{https://doi.org/10.1145/3584931.3607492}
\showDOI{\tempurl}


\bibitem[Shen et~al\mbox{.}(2023a)]%
        {Shen2023ChatGPT}
\bibfield{author}{\bibinfo{person}{Yiqiu Shen}, \bibinfo{person}{Laura Heacock}, \bibinfo{person}{Jonathan Elias}, \bibinfo{person}{Keith~D Hentel}, \bibinfo{person}{Beatriu Reig}, \bibinfo{person}{George Shih}, {and} \bibinfo{person}{Linda Moy}.} \bibinfo{year}{2023}\natexlab{a}.
\newblock \bibinfo{title}{ChatGPT and other large language models are double-edged swords}.
\newblock , \bibinfo{numpages}{e230163}~pages.
\newblock
\urldef\tempurl%
\url{https://doi.org/10.1148/radiol.230163}
\showDOI{\tempurl}


\bibitem[Shi et~al\mbox{.}(2023)]%
        {shi2023badgpt}
\bibfield{author}{\bibinfo{person}{Jiawen Shi}, \bibinfo{person}{Yixin Liu}, \bibinfo{person}{Pan Zhou}, {and} \bibinfo{person}{Lichao Sun}.} \bibinfo{year}{2023}\natexlab{}.
\newblock \showarticletitle{BadGPT: Exploring Security Vulnerabilities of ChatGPT via Backdoor Attacks to InstructGPT}.
\newblock \bibinfo{journal}{\emph{arXiv preprint arXiv:2304.12298}} (\bibinfo{year}{2023}).
\newblock
\urldef\tempurl%
\url{https://doi.org/10.48550/arXiv.2304.12298}
\showDOI{\tempurl}


\bibitem[Shin(2020)]%
        {Shin2020User}
\bibfield{author}{\bibinfo{person}{Donghee Shin}.} \bibinfo{year}{2020}\natexlab{}.
\newblock \showarticletitle{User perceptions of algorithmic decisions in the personalized AI system: Perceptual evaluation of fairness, accountability, transparency, and explainability}.
\newblock \bibinfo{journal}{\emph{Journal of Broadcasting \& Electronic Media}} \bibinfo{volume}{64}, \bibinfo{number}{4} (\bibinfo{year}{2020}), \bibinfo{pages}{541--565}.
\newblock
\urldef\tempurl%
\url{https://doi.org/10.1080/08838151.2020.1843357}
\showDOI{\tempurl}


\bibitem[Steen-Johnsen and Enjolras(2016)]%
        {steen2016fear}
\bibfield{author}{\bibinfo{person}{Kari Steen-Johnsen} {and} \bibinfo{person}{Bernard Enjolras}.} \bibinfo{year}{2016}\natexlab{}.
\newblock \showarticletitle{The fear of offending: Social norms and freedom of expression}.
\newblock \bibinfo{journal}{\emph{Society}}  \bibinfo{volume}{53} (\bibinfo{year}{2016}), \bibinfo{pages}{352--362}.
\newblock
\urldef\tempurl%
\url{https://doi.org/10.1007/s12115-016-0044-2}
\showDOI{\tempurl}


\bibitem[Tanguy et~al\mbox{.}(2016)]%
        {tanguy2016natural}
\bibfield{author}{\bibinfo{person}{Ludovic Tanguy}, \bibinfo{person}{Nikola Tulechki}, \bibinfo{person}{Assaf Urieli}, \bibinfo{person}{Eric Hermann}, {and} \bibinfo{person}{C{\'e}line Raynal}.} \bibinfo{year}{2016}\natexlab{}.
\newblock \showarticletitle{Natural language processing for aviation safety reports: From classification to interactive analysis}.
\newblock \bibinfo{journal}{\emph{Computers in Industry}}  \bibinfo{volume}{78} (\bibinfo{year}{2016}), \bibinfo{pages}{80--95}.
\newblock


\bibitem[Terzis(2020)]%
        {terzis2020onward}
\bibfield{author}{\bibinfo{person}{Petros Terzis}.} \bibinfo{year}{2020}\natexlab{}.
\newblock \showarticletitle{Onward for the freedom of others: marching beyond the AI ethics}. In \bibinfo{booktitle}{\emph{Proceedings of the 2020 Conference on Fairness, Accountability, and Transparency}}. \bibinfo{pages}{220--229}.
\newblock


\bibitem[Thiebes et~al\mbox{.}(2021)]%
        {Thiebes2020Trustworthy}
\bibfield{author}{\bibinfo{person}{Scott Thiebes}, \bibinfo{person}{Sebastian Lins}, {and} \bibinfo{person}{Ali Sunyaev}.} \bibinfo{year}{2021}\natexlab{}.
\newblock \showarticletitle{Trustworthy artificial intelligence}.
\newblock \bibinfo{journal}{\emph{Electronic Markets}}  \bibinfo{volume}{31} (\bibinfo{year}{2021}), \bibinfo{pages}{447--464}.
\newblock
\urldef\tempurl%
\url{https://link.springer.com/article/10.1007/s12525-020-00441-4}
\showURL{%
\tempurl}


\bibitem[Tolmeijer et~al\mbox{.}(2022)]%
        {tolmeijer2022capable}
\bibfield{author}{\bibinfo{person}{Suzanne Tolmeijer}, \bibinfo{person}{Markus Christen}, \bibinfo{person}{Serhiy Kandul}, \bibinfo{person}{Markus Kneer}, {and} \bibinfo{person}{Abraham Bernstein}.} \bibinfo{year}{2022}\natexlab{}.
\newblock \showarticletitle{Capable but amoral? Comparing AI and human expert collaboration in ethical decision making}.
\newblock  (\bibinfo{year}{2022}), \bibinfo{pages}{1--17}.
\newblock
\urldef\tempurl%
\url{https://doi.org/10.1145/3491102.3517732}
\showDOI{\tempurl}


\bibitem[Varona and Su{\'a}rez(2022)]%
        {Varona2022Discrimination}
\bibfield{author}{\bibinfo{person}{Daniel Varona} {and} \bibinfo{person}{Juan~Luis Su{\'a}rez}.} \bibinfo{year}{2022}\natexlab{}.
\newblock \showarticletitle{Discrimination, bias, fairness, and trustworthy AI}.
\newblock \bibinfo{journal}{\emph{Applied Sciences}} \bibinfo{volume}{12}, \bibinfo{number}{12} (\bibinfo{year}{2022}), \bibinfo{pages}{5826}.
\newblock
\urldef\tempurl%
\url{https://doi.org/10.3390/app12125826}
\showDOI{\tempurl}


\bibitem[Veisi et~al\mbox{.}(2025)]%
        {10.1145/3706599.3720152}
\bibfield{author}{\bibinfo{person}{Omid Veisi}, \bibinfo{person}{Khoshnaz Kazemian}, \bibinfo{person}{Farzaneh Gerami}, \bibinfo{person}{Mahya Mirzaee~Kharghani}, \bibinfo{person}{Sima Amirkhani}, \bibinfo{person}{Delong~K. Du}, \bibinfo{person}{Gunnar Stevens}, {and} \bibinfo{person}{Alexander Boden}.} \bibinfo{year}{2025}\natexlab{}.
\newblock \showarticletitle{User Narrative Study for Dealing with Deceptive Chatbot Scams Aiming to Online Fraud}. In \bibinfo{booktitle}{\emph{Proceedings of the Extended Abstracts of the CHI Conference on Human Factors in Computing Systems}} \emph{(\bibinfo{series}{CHI EA '25})}. \bibinfo{publisher}{Association for Computing Machinery}, \bibinfo{address}{New York, NY, USA}, Article \bibinfo{articleno}{560}, \bibinfo{numpages}{7}~pages.
\newblock
\showISBNx{9798400713958}
\urldef\tempurl%
\url{https://doi.org/10.1145/3706599.3720152}
\showDOI{\tempurl}


\bibitem[Vitak et~al\mbox{.}(2016)]%
        {vitak2016beyond}
\bibfield{author}{\bibinfo{person}{Jessica Vitak}, \bibinfo{person}{Katie Shilton}, {and} \bibinfo{person}{Zahra Ashktorab}.} \bibinfo{year}{2016}\natexlab{}.
\newblock \showarticletitle{Beyond the Belmont principles: Ethical challenges, practices, and beliefs in the online data research community}. In \bibinfo{booktitle}{\emph{Proceedings of the 19th ACM conference on computer-supported cooperative work \& social computing}}. \bibinfo{publisher}{Association for Computing Machinery}, \bibinfo{pages}{941--953}.
\newblock


\bibitem[Vitak et~al\mbox{.}(2021)]%
        {10.1145/3462204.3481724}
\bibfield{author}{\bibinfo{person}{Jessica Vitak}, \bibinfo{person}{Michael Zimmer}, \bibinfo{person}{Anna Lenhart}, \bibinfo{person}{Sunyup Park}, \bibinfo{person}{Richmond Y.~Wong}, {and} \bibinfo{person}{Yaxing Yao}.} \bibinfo{year}{2021}\natexlab{}.
\newblock \showarticletitle{Designing for Data Awareness: Addressing Privacy and Security Concerns About “Smart” Technologies}. In \bibinfo{booktitle}{\emph{Companion Publication of the 2021 Conference on Computer Supported Cooperative Work and Social Computing}} (Virtual Event, USA) \emph{(\bibinfo{series}{CSCW '21 Companion})}. \bibinfo{publisher}{Association for Computing Machinery}, \bibinfo{address}{New York, NY, USA}, \bibinfo{pages}{364–367}.
\newblock
\showISBNx{9781450384797}
\urldef\tempurl%
\url{https://doi.org/10.1145/3462204.3481724}
\showDOI{\tempurl}


\bibitem[Waelen(2022)]%
        {Rosalie2022Why}
\bibfield{author}{\bibinfo{person}{Rosalie Waelen}.} \bibinfo{year}{2022}\natexlab{}.
\newblock \showarticletitle{Why AI ethics is a critical theory}.
\newblock \bibinfo{journal}{\emph{Philosophy \& Technology}} \bibinfo{volume}{35}, \bibinfo{number}{1} (\bibinfo{year}{2022}), \bibinfo{pages}{9}.
\newblock
\urldef\tempurl%
\url{https://doi.org/10.1007/s13347-022-00507-5}
\showDOI{\tempurl}


\bibitem[Wang et~al\mbox{.}(2023a)]%
        {Wang2023On}
\bibfield{author}{\bibinfo{person}{Jindong Wang}, \bibinfo{person}{Xixu Hu}, \bibinfo{person}{Wenxin Hou}, \bibinfo{person}{Hao Chen}, \bibinfo{person}{Runkai Zheng}, \bibinfo{person}{Yidong Wang}, \bibinfo{person}{Linyi Yang}, \bibinfo{person}{Haojun Huang}, \bibinfo{person}{Wei Ye}, \bibinfo{person}{Xiubo Geng}, {et~al\mbox{.}}} \bibinfo{year}{2023}\natexlab{a}.
\newblock \showarticletitle{On the robustness of chatgpt: An adversarial and out-of-distribution perspective}.
\newblock \bibinfo{journal}{\emph{arXiv preprint arXiv:2302.12095}} (\bibinfo{year}{2023}).
\newblock
\urldef\tempurl%
\url{https://doi.org/arXiv.2302.12095}
\showDOI{\tempurl}


\bibitem[Wang et~al\mbox{.}(2023b)]%
        {Wang2023Large}
\bibfield{author}{\bibinfo{person}{Peiyi Wang}, \bibinfo{person}{Lei Li}, \bibinfo{person}{Liang Chen}, \bibinfo{person}{Dawei Zhu}, \bibinfo{person}{Binghuai Lin}, \bibinfo{person}{Yunbo Cao}, \bibinfo{person}{Qi Liu}, \bibinfo{person}{Tianyu Liu}, {and} \bibinfo{person}{Zhifang Sui}.} \bibinfo{year}{2023}\natexlab{b}.
\newblock \showarticletitle{Large language models are not fair evaluators}.
\newblock \bibinfo{journal}{\emph{arXiv preprint arXiv:2305.17926}} (\bibinfo{year}{2023}).
\newblock
\urldef\tempurl%
\url{https://doi.org/10.48550/arXiv.2305.17926}
\showDOI{\tempurl}


\bibitem[Weidinger et~al\mbox{.}(2021)]%
        {weidinger2021ethical}
\bibfield{author}{\bibinfo{person}{Laura Weidinger}, \bibinfo{person}{John Mellor}, \bibinfo{person}{Maribeth Rauh}, \bibinfo{person}{Conor Griffin}, \bibinfo{person}{Jonathan Uesato}, \bibinfo{person}{Po-Sen Huang}, \bibinfo{person}{Myra Cheng}, \bibinfo{person}{Mia Glaese}, \bibinfo{person}{Borja Balle}, \bibinfo{person}{Atoosa Kasirzadeh}, {et~al\mbox{.}}} \bibinfo{year}{2021}\natexlab{}.
\newblock \showarticletitle{Ethical and social risks of harm from language models}.
\newblock \bibinfo{journal}{\emph{arXiv preprint arXiv:2112.04359}} (\bibinfo{year}{2021}).
\newblock
\urldef\tempurl%
\url{https://doi.org/10.48550/arXiv.2112.04359}
\showDOI{\tempurl}


\bibitem[Wen et~al\mbox{.}(2019)]%
        {wen2019towards}
\bibfield{author}{\bibinfo{person}{Ruchen Wen}, \bibinfo{person}{Ryan~Blake Jackson}, \bibinfo{person}{Tom Williams}, {and} \bibinfo{person}{Qin Zhu}.} \bibinfo{year}{2019}\natexlab{}.
\newblock \showarticletitle{Towards a role ethics approach to command rejection}. In \bibinfo{booktitle}{\emph{HRI Workshop on the Dark Side of Human-Robot Interaction}}.
\newblock


\bibitem[Wong et~al\mbox{.}(2020)]%
        {10.1145/3406865.3418590}
\bibfield{author}{\bibinfo{person}{Richmond~Y. Wong}, \bibinfo{person}{Karen Boyd}, \bibinfo{person}{Jake Metcalf}, {and} \bibinfo{person}{Katie Shilton}.} \bibinfo{year}{2020}\natexlab{}.
\newblock \showarticletitle{Beyond Checklist Approaches to Ethics in Design}. In \bibinfo{booktitle}{\emph{Companion Publication of the 2020 Conference on Computer Supported Cooperative Work and Social Computing}} (Virtual Event, USA) \emph{(\bibinfo{series}{CSCW '20 Companion})}. \bibinfo{publisher}{Association for Computing Machinery}, \bibinfo{address}{New York, NY, USA}, \bibinfo{pages}{511–517}.
\newblock
\showISBNx{9781450380591}
\urldef\tempurl%
\url{https://doi.org/10.1145/3406865.3418590}
\showDOI{\tempurl}


\bibitem[Yi et~al\mbox{.}(2003)]%
        {yi2003privacy}
\bibfield{author}{\bibinfo{person}{Xun Yi}, \bibinfo{person}{Yiming Ye}, \bibinfo{person}{Chee~Kheong Siew}, {and} \bibinfo{person}{Mahbubur~Rahman Syed}.} \bibinfo{year}{2003}\natexlab{}.
\newblock \showarticletitle{Privacy and Authentication for Agent Supported Cooperative Work}.
\newblock \bibinfo{journal}{\emph{Agent Supported Cooperative Work}} (\bibinfo{year}{2003}), \bibinfo{pages}{273--294}.
\newblock


\bibitem[You(2023)]%
        {you2023impact}
\bibfield{author}{\bibinfo{person}{Leyuan You}.} \bibinfo{year}{2023}\natexlab{}.
\newblock \showarticletitle{The impact of social norms of responsibility on corporate social responsibility short title: The impact of social norms of responsibility on corporate social responsibility}.
\newblock \bibinfo{journal}{\emph{Journal of Business Ethics}} (\bibinfo{year}{2023}), \bibinfo{pages}{1--18}.
\newblock
\urldef\tempurl%
\url{https://doi.org/10.1007/s10551-023-05417-w}
\showDOI{\tempurl}


\bibitem[Zhang et~al\mbox{.}(2023)]%
        {Zhang2023Is}
\bibfield{author}{\bibinfo{person}{Jizhi Zhang}, \bibinfo{person}{Keqin Bao}, \bibinfo{person}{Yang Zhang}, \bibinfo{person}{Wenjie Wang}, \bibinfo{person}{Fuli Feng}, {and} \bibinfo{person}{Xiangnan He}.} \bibinfo{year}{2023}\natexlab{}.
\newblock \showarticletitle{Is chatgpt fair for recommendation? evaluating fairness in large language model recommendation}.
\newblock \bibinfo{journal}{\emph{arXiv preprint arXiv:2305.07609}} (\bibinfo{year}{2023}).
\newblock
\urldef\tempurl%
\url{https://doi.org/10.48550/arXiv.2305.07609 Focus to learn more}
\showDOI{\tempurl}


\bibitem[Zhu(2022)]%
        {Zhu2022AI}
\bibfield{author}{\bibinfo{person}{Junhua Zhu}.} \bibinfo{year}{2022}\natexlab{}.
\newblock \showarticletitle{AI ethics with Chinese characteristics? Concerns and preferred solutions in Chinese academia}.
\newblock \bibinfo{journal}{\emph{AI \& society}} (\bibinfo{year}{2022}), \bibinfo{pages}{1--14}.
\newblock
\urldef\tempurl%
\url{https://doi.org/10.1007/s00146-022-01578-w}
\showDOI{\tempurl}


\bibitem[Zhu et~al\mbox{.}(2020)]%
        {zhu2020blame}
\bibfield{author}{\bibinfo{person}{Qin Zhu}, \bibinfo{person}{Tom Williams}, \bibinfo{person}{Blake Jackson}, {and} \bibinfo{person}{Ruchen Wen}.} \bibinfo{year}{2020}\natexlab{}.
\newblock \showarticletitle{Blame-laden moral rebukes and the morally competent robot: A Confucian ethical perspective}.
\newblock \bibinfo{journal}{\emph{Science and Engineering Ethics}} \bibinfo{volume}{26}, \bibinfo{number}{5} (\bibinfo{year}{2020}), \bibinfo{pages}{2511--2526}.
\newblock


\bibitem[Zhuo et~al\mbox{.}(2023)]%
        {Zhuo2023Exploring}
\bibfield{author}{\bibinfo{person}{Terry~Yue Zhuo}, \bibinfo{person}{Yujin Huang}, \bibinfo{person}{Chunyang Chen}, {and} \bibinfo{person}{Zhenchang Xing}.} \bibinfo{year}{2023}\natexlab{}.
\newblock \showarticletitle{Exploring ai ethics of chatgpt: A diagnostic analysis}.
\newblock \bibinfo{journal}{\emph{arXiv preprint arXiv:2301.12867}} (\bibinfo{year}{2023}).
\newblock
\urldef\tempurl%
\url{https://www.arxiv-vanity.com/papers/2301.12867/}
\showURL{%
\tempurl}


\bibitem[Zoshak and Dew(2021)]%
        {zoshak2021beyond}
\bibfield{author}{\bibinfo{person}{John Zoshak} {and} \bibinfo{person}{Kristin Dew}.} \bibinfo{year}{2021}\natexlab{}.
\newblock \showarticletitle{Beyond kant and bentham: How ethical theories are being used in artificial moral agents}.
\newblock  (\bibinfo{year}{2021}), \bibinfo{pages}{1--15}.
\newblock
\urldef\tempurl%
\url{https://doi.org/10.1145/3411764.3445102}
\showDOI{\tempurl}


\end{thebibliography}

\onecolumn % Switch to one-column mode to allow longtable

\appendix % Start the appendix section
\section{Appendix: Questionnaire Results}

\renewcommand{\thetable}{A\arabic{table}} % Custom table numbering for appendix
\setcounter{table}{0}

\begin{longtable}{p{0.6cm}p{9cm}p{1.2cm}p{1cm}p{1cm}p{1cm}p{1cm}}
\caption{The online questionnaire of the Likert scale} \label{Table2} \\
\toprule
ID & Question & Strongly disagree & Disagree & Neutral & Agree & Strongly agree \\
\midrule
\endfirsthead

\toprule
ID & Question & Strongly disagree & Disagree & Neutral & Agree & Strongly agree \\
\midrule
\endhead

\midrule
\endfoot

\bottomrule
\endlastfoot

\multicolumn{7}{l}{\textbf{Bias}} \\
\hline
Q1 & Local communities should be role models for ChatGPT to respond to user questions (e.g., when ChatGPT responds, a person from Iran should consider the roles, laws, and social norms in Iran). & 8 & 23 & 36 & 25 & 6 \\
Q2 & ChatGPT seems to show bias when asked sensitive political questions(e.g. When German politicians ask specific questions about policies or the history of wars, ChatGPT responds in a targeted manner). & 13 & 25 & 49 & 15 & 6 \\
\addlinespace[0.5em]
\hline
\multicolumn{7}{l}{\textbf{Trustworthiness}} \\
\hline
Q3 & ChatGPT's actions as ethically sound (e.g., if you ask an unethical prompt, ChatGPT will not react or provide a warning.) & 13 & 28 & 23 & 37 & 9 \\
Q4 & ChatGPT can understand the social reality and user interaction (e.g., if an Iranian person asks a question about New Year's, the response differs compared to German or US people.) & 14 & 34 & 26 & 28 & 6 \\
Q5 & Using ChatGPT can bring some security risks for users (e.g., users that share data with ChatGPT 3.5 accept the risk of leaking some of them). & 13 & 45 & 17 & 31 & 4 \\
Q6 & Understanding how ChatGPT's results were generated is somewhat challenging for an outside observer. & 29 & 42 & 24 & 11 & 3 \\
Q7 & ChatGPT's output is reliable and understandable for the user (e.g., a prompt about an environmental issue generates a valid and acceptable response). & 9 & 32 & 41 & 25 & 1 \\
\addlinespace[0.5em]
\hline
\multicolumn{7}{l}{\textbf{Security}} \\
\hline
Q8 & ChatGPT can be targeted by hackers and used for hacking (e.g., a hacker can use ChatGPT to develop a virus or Trojan for stealing information from a website of a personal computer). & 13 & 39 & 24 & 24 & 8 \\
Q9 & ChatGPT has the potential to be used in online attacks (e.g., scammers can use ChatGPT for online fraud). & 25 & 25 & 30 & 18 & 10 \\
Q10 & ChatGPT has effectively ensured the safety and privacy of the user's identity and data (e.g., your information shared as an input (prompt) is not accessible outside of ChatGPT). & 14 & 33 & 33 & 22 & 6 \\
Q11 & ChatGPT have semantic consistency in similar prompts. (Mean; if you write two prompts with the same mean but semantically different, the result will be the same for both of the prompts.). & 19 & 48 & 24 & 13 & 5 \\

\addlinespace[0.5em]
\hline
\multicolumn{7}{l}{\textbf{Toxicity}} \\
\hline
Q12 & We should limit ChatGPT and establish specific boundaries for developments and responses. (e.g., ChatGPT should be limited to responding to users in the roles of a doctor or a psychologist.) & 10 & 19 & 11 & 45 & 24 \\
Q13 & ChatGPT should prioritize nonhuman interests over human interests. (e.g., decision-making about saving animal and plant species during city development compared to cost and economic issues). & 48 & 35 & 17 & 7 & 2 \\
Q14 & The ChatGPT output is transparent and trustworthy for the user. (e.g., the user can find the precise response they seek on ChatGPT). & 4 & 29 & 26 & 42 & 8 \\
Q15 & ChatGPT can be used to detect hate speech in the input context and user prompts (e.g., if the user tells ChatGPT an insulting word, ChatGPT can ignore it). & 14 & 39 & 14 & 32 & 11 \\
Q16 & ChatGPT does not have permission to write about hate speech (e.g., if users ask for a sexual harassment story, ChatGPT shows them a warning about that.) & 8 & 20 & 37 & 40 & 4 \\
\addlinespace[0.5em]
\hline
\multicolumn{7}{l}{\textbf{Social Norms}} \\
\hline
Q17 & ChatGPT's output based on different social norms is fair in different countries (e.g., ChatGPT responds to questions about New Year celebrations based on user location). & 12 & 25 & 38 & 24 & 10 \\
Q18 & Consequences using ChatGPT with users working with them. (e.g., if the user uses them as a doctor, they should be aware of the risk of advising.). & 5 & 19 & 21 & 58 & 7 \\
Q19 & Using ChatGPT impacts our autonomy and freedom (e.g., ChatGPT has the potential to influence and manipulate our decision-making when we are considering a presidential candidate.) & 13 & 26 & 23 & 27 & 21 \\
Q20 & The human factor impacts interactions and decisions made by ChatGPT (e.g., the user can consider whether the output is accurate or needs more information and investigation). & 9 & 32 & 41 & 26 & 1 \\
\addlinespace[0.5em]
\hline
\multicolumn{7}{l}{\textbf{Ethics}} \\
\hline
Q21 & The approaches for collecting data should be based on ethical approaches (e.g., the OpenAI company should get permission from the user when they want to collect data). & 10 & 22 & 42 & 28 & 7 \\
Q22 & ChatGPT output is ethical for users generally. (e.g., ChatGPT responds to users based on ethical codes from the company, such as avoiding hurting users by giving advice). & 27 & 33 & 31 & 16 & 2 \\

\end{longtable}


\end{document}